\documentclass[aps,prb,twocolumn,showpacs]{revtex4}
\usepackage{boxedminipage}
\usepackage{lscape}
\usepackage{float}
\usepackage[ansinew]{inputenc}
\usepackage{epsfig}
\usepackage{subfigure}
\usepackage{graphicx}
\usepackage{amsmath}
\usepackage{amssymb}
\usepackage{amsthm,url}
\usepackage[french,english]{babel}  
\usepackage{epstopdf}
\usepackage{amsfonts}
\renewcommand{\d}[2]{\frac{#1}{#2}}
\newcommand{\pd}{\partial}

\newcommand{\barray}{\begin{eqnarray}}
\newcommand{\earray}{\end{eqnarray}}

\newcommand{\beq}{\begin{equation}}

\newcommand{\eeq}{\end{equation}}

\begin{document}
\selectlanguage{english}

\title{Machine learning for many-Body physics: The case of the Anderson impurity model}
\author{Louis-Fran\c{c}ois Arsenault$^{1}$, Alejandro Lopez-Bezanilla$^{2}$, O. Anatole von Lilienfeld$^{3,4}$, and Andrew J. Millis$^{1}$}
\affiliation{$^1$ Department of Physics, Columbia University, New York, New York 10027, USA\\
$^{2}$ Materials Science Division, Argonne National Laboratory, 9700 S. Cass Avenue, Lemont, IL 60439, USA\\
$^{3}$ Institute of Physical Chemistry, Department of Chemistry, University of Basel, Klingelbergstrasse 80, CH-4056 Basel, Switzerland\\
$^{4}$ Argonne Leadership Computing Facility, Argonne National Laboratory, 9700 S. Cass Avenue, Lemont, IL 60439, USA}

\date{\today}

\begin{abstract}
Machine learning methods are applied to finding the Green's function of the Anderson impurity model, a basic model system of quantum many-body condensed-matter physics. Different methods of parametrizing the Green's function are investigated; a representation in terms of Legendre polynomials is found to be superior due to its limited number of coefficients and its applicability to state of the art methods of solution. The dependence of the errors on the size of the training set is determined. The results indicate that a machine learning approach to dynamical mean-field theory may be feasible.
\end{abstract}


\maketitle

\hyphenation{Brillouin}

\section{Introduction}
The fermionic quantum many-body problem is believed to be in the class of problems whose full solution is exponentially hard \cite{TroyerWiese2005}. Approximate methods have been developed, but many of these are also extremely computationally demanding. There is, therefore, an ongoing search for efficient approximate methods, useful, for example, in computational surveys of wide classes of materials, or to provide a first look at a complicated situation.\\
\\
The predominant approach has been to use simplifying approximations, for example, truncated perturbation-theory series expansions, variants of mean-field theory, quasiclassical approximations, or analytical interpolation functions. The development of machine learning (ML) techniques in computer science motivates us to explore a complementary approach. ML provides an estimate of the result of a calculation based on interpolation from a statistical analysis of datasets of solved problems \cite{MLbook}. ML is  widely used in many big-data applications, and has recently been proposed as a method for obtaining approximate solutions of the equations of density functional theory (DFT)\cite{ML4Kieron2012}, of the molecular electronic Schr\"odinger equation~\cite{RuppPRL2012,Montavon2013}, and of transmission coefficients for electron quantum transport~\cite{QuantumTransportML2014}. ML is also used to construct force-fields from molecular dynamics\cite{SumpterNoidNeuralNetworks1992,Neuralnetworks_Scheffler2004,Manzhos2006,Neuralnetworks_BehlerParrinello2007,bpkc2010}.\\
\\
In this paper we investigate ML techniques to infer solutions to the quantum many-body problem arising in applications of the dynamical mean-field theory (DMFT) method.\cite{dmft1,Gaby_Antoine_DMFT,Jarrell_DMFT} DMFT has become widely used in condensed-matter physics and materials science for obtaining nonperturbative information about materials with strong local correlations. While DMFT is an approximation to the full many-body problem, it does require the solution of a fully interacting quantum impurity model (a quantum field theory defined in  zero space but one time dimension), and accurate solutions require substantial numerical effort which is time consuming even with modern algorithms and hardware\cite{Werner_CTQMC,Haule_CTQMC,Gull_review}. A sufficiently accurate ML model of DMFT could provide an inexpensive \emph{solver}, useful for rapid preliminary screening of wide ranges of materials and as a method for identifying promising starting points for further refinement using more expensive and sophisticated methods or experiments.
\\
\\
In its conventional formulation, DMFT maps one function of frequency into another. The input is the \emph{bare hybridization function}, which encapsulates  relevant information about the  crystal structure and quantum chemistry of a material via a representation of what  the local density of electronic excitations would be if many-body correlations were neglected. A small number of additional parameters, such as the on-site interaction strength, must also be specified. The output is the electron Green's function (or equivalently self-energy), which provides an approximation to the exact density of states (DOS) obtained by including local effects of many-body correlations.\\
\\
Implementing a ML approach to DMFT thus entails constructing a training set of physically  reasonable hybridization functions, determining the spectral functions corresponding to the training examples, and constructing a model that provides the needed interpolation formula. Such a ML procedure goes beyond previous applications of ML to electronic structure because we are mapping a function to a function whereas the ML approaches to DFT or the Schr$\ddot{\text{o}}$edinger equation provide only a small number of scalar outputs, such as the total energy of atomization, ionization potential, or excitation energy\cite{Montavon2013,Hansen2013}. A key issue is thus to devise an efficient representation of the functions of interest in terms of a reasonably sized set of parameters. The first application we foresee for real systems is a material science computation tool to optimize a desired property.\\
\\
In this paper we address this key issue for the Anderson impurity model (AIM),  the archetypical quantum impurity model.  For this model the input hybridization function is known {\em a priori}, and can be specified by few parameters. The focus therefore lies on the prediction of the output, namely, the electron Green's function. In future work we will discuss ML applications to the full DMFT problem of determining the self-consistent relation between the Green's function and hybridization function, starting  from an arbitrary hybridization function.
\\
\\
This paper is organized as follows: in Sec.~\ref{ML_supervised}, we summarize the supervised learning approach. In Sec.~\ref{ML_supervised_repres} we discuss how to represent a function for ML and in Sec.~\ref{ML_supervised_Kernel} we discuss the kernel ridge regression that we use in this paper, while in Sec.~\ref{Quality_ML} we show how the ML procedure is tested. In Sec.~\ref{QI_AIM} the one site AIM and its solution by exact diagonalization are presented. In Sec.~\ref{QI_ML_AIM} we present the machine learning solution. Section~\ref{QI_metho} discusses the methodology of the calculation. Section~\ref{QI_repre_G} presents in detail the four types of representation for the Green's functions  we study: in Sec.~\ref{QI_repre_G_cont_frac} the continued fraction, in Sec.~\ref{QI_sec_Gwn} the Matsubara frequency, in Sec.~\ref{QI_sec_Gtau} imaginary time, and in Sec.~\ref{QI_sec_Gl} Legendre polynomials.  The results for these different representations are shown in Secs.~\ref{QI_sec_cont_frac_results}, \ref{QI_sec_Gwn_results}, \ref{QI_sec_Gtau_results} and \ref{QI_sec_Gl_results}. In Secs.~\ref{QI_sec_prediction_size} and \ref{QI_sec_mass_enh} we show how the size of the learning set affects the predictions of the DOS and the mass enhancement. Finally, in Sec.~\ref{QI_predic_min_set} we look at how we can define an absolute minimal learning set for our problem. Section~\ref{sum_concl} is the summary and conclusion. Appendix~\ref{ML_appen} gives details on the kernel ridge regression, Appendix~\ref{ED_appen} gives details of the exact diagonalization method, Appendix~\ref{Legendre_appen} gives details of the representation of  the Green's function using Legendre polynomials, and Appendix~\ref{Appen:eff_alpha} presents the derivation of the effective $\alpha$ matrix for the Legendre polynomials approach.
\section{Supervised learning for a function}\label{ML_supervised}
\subsection{Representation of the function}\label{ML_supervised_repres}
We wish to learn a function of one complex variable, $f(z)$, with $f$ representing the Green's function or self-energy. The model whose solution gives $f(z)$ is specified by a descriptor, $\mathbf{D}$, the set of input parameters needed to describe the model.  Where appropriate we denote the dependence of $f$ on the input parameters as $f(\mathbf{D};z)$. The ML approach is to infer $f(\mathbf{D};z)$ from a given set of $N_L$ results $f(\mathbf{D}_{l=1...N_L};z)$.\\
\\
The functions of physical interest have a spectral representation
\begin{equation}\label{spectral}
f(z)=\int \frac{dx}{\pi}\frac{f''(x)}{z-x},
\end{equation}
with $x$ a real number. The spectral function  $f''(x)$ is non-negative, integrable, and typically nonzero only over a finite range of $x$.\\
\\
While $f$ is fully specified by $f''$, for technical reasons one often has data only for, or is only interested in, $f$ on the \emph{Matsubara frequencies} $z=i\omega_n=i\left(2n+1\right)\pi T$ where $T$ is the temperature, $i = \sqrt{-1}$ and $n$ is an integer. $f$ is sometimes also studied on   the \emph{imaginary time} ($\tau$) axis in the interval $0 < \tau < \beta=1/T$. The $\tau$-dependent function is related to the Matsubara frequency values by
\begin{equation}
f(\tau)=T\sum_{\omega_n}f\left(i\omega_n\right)e^{-i\omega_n\tau}=-\int \frac{dx}{\pi}   f''(x)\d{\text{e}^{-x\tau}}{1+\text{e}^{-\beta x}}.
\label{ftau}
\end{equation}
It is worth noting that the definition of the Matsubara frequencies along with Eq.~\eqref{ftau} implies three important results that will be used later:
\begin{equation}\label{anti_tau}
\begin{split}
    f(\tau-\beta)&=-f(\tau)\\
    f(\beta^-)+1 &= - f(0^+)\\
    \d{df(\tau)}{d\tau}\Big|_{\tau = \beta^-} &= \int \d{dx}{\pi}xf''(x) - \d{df(\tau)}{d\tau}\Big|_{\tau = 0^+},
\end{split}
\end{equation}
where $\int \d{dx}{\pi}xf''(x)$ is the first moment of the spectral function which  depends upon the particular Hamiltonian.\\
\\
Equation~\eqref{spectral} implies that the values of $f(z)$ at different $z$ values are correlated. This presents a challenge since the machine learning algorithms we will use treat the different components of $f$ (or the coefficients of its expansion) as independent. Therefore, by independently adding numerical errors on each point the machine may not fully respect the needed correlations and constraints. Put differently, the $f$ predicted by our machine learning algorithm could arise from a spectral function with negative regions. This issue is not unique to ML and is related to the well-known \emph{analytical continuation} problem of inverting Eq.~\eqref{spectral} to determine  $f''(x)$ from the full set of values of $f(i\omega_n)$. While general theorems from complex variable theory imply that the inversion is possible in principle, the kernel $1/(z-x)$ of Eq.~\eqref{spectral} has many very small eigenvalues so the inversion problem is ill conditioned. However, in all the cases we have considered, these issues seem not to pose any difficulties in practice, so we proceed straightforwardly, with the assumptions justified {\em a posteriori}.\\
\\
In any numerical approach to problems involving continuous functions, a choice of discretization must be made. Any discretization expresses the continuous function $f(z)$ in terms of a set of $N$  numbers $f_{m=1...N}$ which we assemble into an $N$-dimensional vector $\boldsymbol{f}$:
\begin{equation}\label{function_ML}
  f(\mathbf{D};z)\rightarrow \boldsymbol{f}(\mathbf{D}) = \{f_1(\mathbf{D}),f_2(\mathbf{D}),\ldots,f_N(\mathbf{D})\}.
\end{equation}
One must note that the $f_m$'s need not be direct points of the function but can also be coefficients that define the function.\\
\\
We have considered four discretizations of $f(z)$: (1) a continued fraction representation; (2) values on the Matsubara axis $f(i\omega_n)$ up to some cutoff frequency $|\omega_n|<\Omega_c$; (3) discrete  values on the imaginary time axis, $f_j=f(j\Delta \tau/T)$; and (4) an expansion of $f(\tau)$ in terms of orthogonal (Legendre) polynomials. Both the continued fraction and Legendre polynomials representation lead to a correctly normalized and non-negative spectral function, but as will be seen the continued fraction representation is for other reasons not optimal. The other representations do not necessarily lead to a $f(z)$ which has a spectral representation with a non-negative spectral function for small and random sets of $N_L$ results, but, as will be seen, they seem to work well in practice.
\subsection{Machine learning: kernel approach}\label{ML_supervised_Kernel}
In this paper we apply the kernel approach\cite{Kernel_ridge}, which involves two ingredients: a \emph{distance kernel} $K_m(\mathbf{D}_l,\mathbf{D})$, a symmetric and positive definite function, fixed {\em a priori}, and a coefficient matrix $\alpha_{lm}$ which is to be determined. In terms of these, one generates an approximation $g$ to the desired $f$  given by the so-called kernel ridge regression (KRR), an expansion in an abstract kernel space. In terms of the expansion defined in Eq.~\eqref{function_ML}) we approximate the components $f_m$ of $\boldsymbol{f}$ in terms of approximate components $g_m$ given by
\begin{equation}
f_m(\mathbf{D})\approx g_m(\mathbf{D})=\sum_l\alpha_{lm}K_m(\mathbf{D}_l,\mathbf{D}).
\label{MLfundamental}
\end{equation}
Here we use the subscript $m$ to label the entries in $f$ (i.e., the different components in the discrete representation of the function: direct points or appropriate coefficients) and $l$ to label the training examples.\\
\\
The coefficients $\alpha_{lm}$ are calculated by minimizing a cost function defined in terms of the difference between the approximations $g_m(\mathbf{D}_l )$ and the exactly known $f_m(\mathbf{D}_l)$, with the addition of a set of  Lagrange multipliers introduced to regularize the problem.  In our approach we assume that each component $m$ of $\boldsymbol{f}$ is learned separately and independently, so we define a cost function separately for each $m$  as
\begin{eqnarray}\label{cost_set}
  C_m &=& \sum_l\left( g_m(\mathbf{D}_l) - f_m(\mathbf{D}_l) \right)^2
  \\
  &&+ \lambda_m\sum_{l,p}\alpha_{lm}K_m(\mathbf{D}_l,\mathbf{D}_p)\alpha_{pm}.
  \nonumber
\end{eqnarray}
Minimization then gives at fixed $m$ (see Appendix~\ref{ML_appen} for details)
\begin{equation}\label{alphadef}
\boldsymbol{\alpha}_m = \left( \overline{\overline{ \boldsymbol{K}}}_m + \lambda_m\overline{\overline{ \boldsymbol{I}}} \right)^{-1}\boldsymbol{f}_m.
\end{equation}
In Eq.~\eqref{alphadef}, $\boldsymbol{\alpha}_m$ is a column vector of length $N_L$ containing the different $\alpha$ for $m$ fixed and $\boldsymbol{f}_m$ is also a column vector of length $N_L$ but containing the different values of $f_m$ ($m$ fixed) on the training set ($\mathbf{D}_{l=1\ldots N_L}$). It is important to note that this vector $\boldsymbol{f}_m$ is not the same as the vector $\boldsymbol{f}$ in Eq.~\eqref{function_ML} which is a vector of length $N$ containing the values of $f_{m=1\ldots N}$ for one specific $\mathbf{D}$. Finally, $\overline{\overline{ \boldsymbol{K}}}_m$ and $\overline{\overline{ \boldsymbol{I}}}$ are the Kernel and identity matrices of size $N_L\times N_L$.\\
\\
In our actual calculations we assume that the kernel $K$ and the Lagrange multiplier $\lambda$ are independent of  $m$, i.e., the same for all components  of the function to be learned. We may then assemble all of the examples $\boldsymbol{f}_m$ into a $N_L\times N$ matrix $\overline{\overline{ \boldsymbol{f}}}$ in which each column is for a different $m$ and combine Eq.~\eqref{alphadef} into a unique matrix equation
\begin{equation}\label{alpha_mat_Kunique}
\overline{\overline{ \boldsymbol{\alpha}}}=\left(\overline{\overline{ \boldsymbol{K}}}+\lambda\overline{\overline{ \boldsymbol{I}}}\right)^{-1}\overline{\overline{ \boldsymbol{f}}}.
\end{equation}
This makes the process of learning a function very efficient since even if $\boldsymbol{\overline{\overline f}}$ is a very large matrix, the solution needs to be obtained only once. For $K$, we use the weighted exponential kernel
 \begin{equation}\label{Kernel_def}
    K(\mathbf{D}_i,\mathbf{D}) = \text{e}^{-\d{|\boldsymbol{d}_i|}{\sigma}},
  \end{equation}
where $|\boldsymbol{d}_l| = |\text{D}_{l1}-\text{D}_{1}|+ |\text{D}_{l2}-\text{D}_{2}| + \ldots$ is the Manhattan distance between the two parameter sets and $\sigma$ gives the radius of effect that a particular point of the data set $\mathbf{D}_l$ will have in the prediction process\cite{Hansen2013}.\\
\\
For numerical calculation, direct matrix inversion should be avoided and Eq.~\eqref{alpha_mat_Kunique} is solved in the form $\left(\boldsymbol{\overline{\overline K}}+\lambda\textbf{I}\right)\overline{\overline{ \boldsymbol{\alpha}}} = \boldsymbol{\overline{\overline f}}$. Since $\boldsymbol{\overline{\overline K}}$ is a square, real, symmetric, and positive definite matrix, a standard Cholesky solver can be used. The process is very fast; for example, on a desktop computer with a Intel Core I7-4770 quad core at 3.40 GHz, using a dataset of 5000 points, it takes about 4 s to learn 2400 parameters defining $G$ in its continued fraction representation plus the ground-state energy.\\
\\
The way we \emph{train} our machine is to choose $N_L$ results from a database of solutions. This subset of solutions is called the training set and we keep the remaining solutions in the database apart and consider a subset of them as the test set. The training set is then used to construct $\overline{\overline{ \boldsymbol{K}}}$ and Eq.~\eqref{alpha_mat_Kunique} can be solved. We then use the test set to check if the predictions from the ML give accurate results and thus can be used to predict solutions not contained in the database. More details on the training process we use are given in Sec.~\ref{QI_metho}.\\
\\
As shown in this section, we can see that ML is an interpolation approach, never an extrapolation. The obvious question of how small the shift in parameters of a desired prediction $\mathbf{D}$ from the $\mathbf{D}_i's$ must be (i) depends on the training set density (the $N_L$ examples that will be used to obtain the $\alpha_{lm}$), (ii) depends on it's distribution, and (iii) still has to be investigated in more rigorous ways than is done either this paper or in the literature in general.
\subsection{Quality of machine learning}\label{Quality_ML}
In most ML studies, the quality of the machine is assessed in terms of the mean absolute error (MAE) of the predicted quantities~\cite{RuppPRL2012,Montavon2013}. Because we are interested in predicting a function we introduce a different metric, the average relative difference (ARD), defined as
\begin{equation}\label{ARD}
  ARD = 100\left\langle  \d{\left|f_{predic}(z)-f_{exact}(z)\right|}{\left|f_{exact}(z)\right|} \right\rangle_z,
\end{equation}
where the factor $100$ is introduced so that the ARD is given in percentage and $\langle\rangle_z$ denotes the average over a suitably defined set of function arguments $\{z_i\}$. In practice we take the set of arguments $z_i$ to be the first few Matsubara frequencies corresponding to energies up to the typical scales of the problem; the values at higher $\omega_n$ are typically controlled by sum rules and are small, with very small errors. The Matsubara frequencies are an appropriate measure because (i) they contain the relevant physical information, (ii) there is a natural discretization, and (iii) the use of a common estimator permits straightforward comparison of different approaches. Note that three of the four representations (the continued fraction, the Matsubara frequency, and Legendre polynomials) give $G(i\omega_n)$ in a direct way. We only compare $\text{Im}\{G(i\omega_n)\}$.\\
\\
However, the ARD or any measure that give an error representation for the entire function as a single number can be misleading, because it can mask narrow but important regions of the function which might be badly predicted. Therefore, as a further test of the quality of the machine, we examine in detail, for two specific cases the predictions of the ML for (a) $\text{Im}\{G(i\omega_n)\}$ (b) the density of states obtained by using Pad\'{e} analytical continuation, and (c) the low-frequency renormalization factor $Z = \left(1-\d{\pd \text{Re}\left\{\Sigma\right\}}{\pd\omega}\Big|_{\omega\rightarrow 0}\right)^{-1}$ estimated as the extrapolation
\begin{equation}\label{mass_enh}
  Z = \d{\text{Im}\left\{\Sigma(i\omega_n)\right\}}{\omega_n}\Big|_{\omega_n\rightarrow 0}.
\end{equation}
This renormalization factor is a unique number characterizing an important low-frequency property of the model.
\section{Quantum Impurity Model}\label{QI_AIM}
\begin{figure}[tpb]
  \begin{center}
  \includegraphics[scale=0.35]{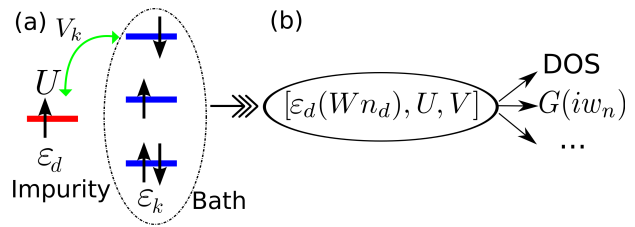}
  \end{center}
  \caption{(Color online) Representation of the Anderson impurity model (AIM): (a) An impurity with zero, one or two localized electrons hybridizes with a partially filled bath of non interacting electrons (b) With fixed bath DOS and constant hybridization $V$, the AIM can be uniquely described by three parameters $[\varepsilon_d (Wn_d),U,V]$ and be used in a ML approach to predict correlation functions and other physical properties.}
  \label{fig:AIM}
\end{figure}

In this paper we apply machine learning techniques to the  single impurity Anderson impurity model.  The Hamiltonian is
\begin{equation}\label{H_AIM}
\begin{split}
  H = &\sum_{\sigma}\varepsilon_dd_{\sigma}^{\dagger}d_{\sigma} + Ud_{\uparrow}^{\dagger}d_{\uparrow}d_{\downarrow}^{\dagger}d_{\downarrow}+ \sum_{k,\sigma}\varepsilon_k c_{k\sigma}^{\dagger}c_{k\sigma}\\ &+ \sum_{k,\sigma}V_k\left( d_{\sigma}^{\dagger}c_{k\sigma} + c_{k\sigma}^{\dagger}d_{\sigma} \right).
\end{split}
\end{equation}
As shown in Fig~\ref{fig:AIM}(a), this represents one localized level (with two-fold spin degeneracy), embedded in a bath of noninteracting electrons. The localized electronic states are represented by the creation (annihilation) operators $d_{\sigma}^{\dagger} (d_{\sigma})$ ($\sigma$ the spin) while the continuous bath states are represented by the operators $c_{k\sigma}^{\dagger} (c_{k\sigma})$. The on-site energy of the impurity is $\varepsilon_d$ and the interaction term which penalizes double occupancy if positive, is $U$. The bath dispersion is $\varepsilon_k$ and the hybridization between the bath and the impurity is $V_k$, where $k$ is a wave number.
\\
\\
The relevant features of the bath and hybridization are encoded by the hybridization function $\Delta$ which for this model is
\begin{equation}
\Delta(z)=\sum_k\frac{|V_k|^2}{z-\varepsilon_k}
\label{Deltadef}
\end{equation}
We choose the bath to have a semicircular density of states and to define the energy unit to be the half width $W = 1$, so
\begin{equation}
N_0(\varepsilon) = \d{2}{\pi}\Theta (1-|\varepsilon|)\sqrt{1-\varepsilon^2}.
\label{BathDOS}
\end{equation}
We further choose the hybridization to be a constant, $V$ and set the chemical potential to be equal to the energy at the center of the bath density of states $\mu=0$ in the notations of Eq.~\eqref{BathDOS}. The parameters of the model are then $U,V$, and $\varepsilon_d$. We find it more convenient to replace the on-site energy  $\varepsilon_d$ by the occupation $n_d=\langle d^\dagger_\uparrow d_\uparrow+d^\dagger_\downarrow d_\downarrow\rangle$, which is a single-valued function of $\varepsilon_d$ for nonzero $V$ and which we multiply by the half bandwidth $W$ so it has the same dimension as the other parameters. The descriptor will be a vector of three values
\begin{equation}\label{descriptor}
  \mathbf{D} = [U,V,Wn_d].
\end{equation}
The machine learning task is then to predict the Green's function of the model, $G_\sigma(\tau)=-\langle Tr_{\tau} d_\sigma(\tau)d^\dagger_\sigma(0)\rangle$, in terms of $(U,V,n_d)$. This ML representation of the AIM, as well as what the prediction leads to, is shown in Fig~\ref{fig:AIM}(b).\\
\\
The Anderson impurity model is a full many-body problem, and while an exact Bethe-ansatz solution exists for particular choices of parameters\cite{AIM_Bethe_Ant_Wiegmann}, there is no known exact analytic solution  for the general case. While many different methods are available for solving the problem including approximate ones (such as the renormalized strong-coupling expansion proposed by Krivenko \emph{et al.} [\onlinecite{Krivenko}] and diagrammatic resummation methods such as the non-Crossing Approximation (NCA) and one-crossing approximation (OCA) as in [\onlinecite{Kirchner}]) and numerically exact ones (quantum Monte Carlo (QMC) [\onlinecite{HirschFye86,Werner_CTQMC,Gull_review,WangMillis2008}] and numerical renormalization group (NRG) [\onlinecite{Bulla_NRG_review}]), we will use the exact diagonalization (ED) method [\onlinecite{dmft2,Dagotto}], in which the hybridization function is represented as a sum of small number ($N_b$) poles and weights in the form
\begin{equation}\label{ED_hyb}
  \Delta (z) \approx \Delta_{N_b} = \sum_{l=1}^{N_b}\d{V_l^2}{z-\varepsilon_l}.
\end{equation}
One is then left with a finite-size Hamiltonian (size $N_s = N_b + 1$) which can be diagonalized exactly, allowing computation of the many-body ground-state wave function and energy. The Green's function is obtained as a continued fraction:
\begin{equation}\label{G_cont_frac}
\begin{split}
  G_{\sigma}(z) = &\d{\langle GS|d_{\sigma}d_{\sigma}^{\dagger}|GS\rangle}{z+E_{GS}-a_0^>-\d{b_1^{>2}}{z+E_{GS}-a_1^>-\d{b_2^{>2}}{z+E_{GS}-a_2^> - \ddots}}}\\ &+ \d{\langle GS|d_{\sigma}^{\dagger}d_{\sigma}|GS\rangle}{z-E_{GS}-a_0^<-\d{b_1^{<2}}{z-E_{GS}-a_1^<-\d{b_2^{<2}}{z-E_{GS}-a_2^< - \ddots}}}.
\end{split}
\end{equation}
We define the exact Green's function as Eq.~\eqref{G_cont_frac} truncated at 600 continued fraction coefficients but note that the results for our model are not materially different if only the first 100 coefficients are retained.  We use Eq.~\eqref{G_cont_frac} to compute $G$ on the real axis, on the Matsubara axis (with a fictitious temperature $\beta = 1/T = 200$), or as a function of imaginary time by Fourier transform. Appendix~\ref{ED_appen} gives more details about ED.

\section{Machine learning for the AIM}\label{QI_ML_AIM}

\subsection{Methodology of the calculations}\label{QI_metho}
We created a database of examples as input to the machine learning process by solving the Anderson impurity model for $5000$ combinations of $U$,$V$, and $n_d$.  The interaction $U$ was varied from a small value to twice the bandwidth 0.16 to 4 in 25 equal intervals. The hybridization $V$ is varied from a small value to the order of half the bandwidth 0.1 to 0.75 in 20 equal intervals. Finally, the filling of the impurity $n_d$ is varied from 0.6 to 1.4 (0-40\% doping in both sides) in intervals of 0.1 and we also included $n_d=0.95$. For ED, we have used $N_b= 7$ and as already mentioned a fictitious temperature of $\beta = 1/T = 200$. For the AIM with a bath with finite bandwidth, in the low-temperature limit, the Kondo temperature is given by\cite{Haldane1978}
\begin{equation}\label{TK}
  T_K \approx 0.2\sqrt{2\Gamma U}\text{e}^{\d{\pi\varepsilon_d(\varepsilon_d+U)}{2\Gamma U}},
\end{equation}
where $\Gamma \equiv \Delta (0) = 2V^2$. The validity of Eq.~\eqref{TK} is limited for $\varepsilon_d(\varepsilon_d+U) < 0$. This leads for our parameters to a $T_K$ in the range from about $6\text{x}10^{-36}$ to a maximum of about 0.48. This maximal $T_K$ is obtained for maximum $U$, $V$, and doping, i.e., $U = 4$, $V = 0.75$, and $n_d = 0.6$ or $1.4$.\\
\\
From the database of 5000 solutions we randomly choose $N_{ML}$  results to serve as  the learning and the test sets. These $N_{ML}$ results are then divided into   $M$ subsets.  One is the testing set and the remaining $M-1$ form the learning set. For example, to calculate the ARD, we always use test sets of 100 such that we will choose $M = N_{ML}/100$. We use a null Lagrange multiplier ($\lambda = 0$) and $\sigma$ is most of the time equal to $\sigma = 10W$ except for one case where we considered $\sigma = 1W$.
\\
\\
\subsection{Representations of the Green's function}\label{QI_repre_G}
We have investigated four different representations of  the Green's function: (i) the continued fraction representation, Eq.~\eqref{G_cont_frac}; (ii) the Green's function in Matsubara frequency; (iii) the  Green's function in imaginary time; and finally, (iv) a representation of the imaginary time Green's function as a sum of Legendre polynomials. 
\subsubsection{Continued fraction representation}\label{QI_repre_G_cont_frac}
Here we proceed directly from Eq.~\eqref{G_cont_frac}. We must learn some number $N_c$ of coefficients $a$ and $b$ for the particle and hole Green's function, along with the ground-state energy, thus $4N_c+1$ coefficients.
Hence $\boldsymbol{f}$ (Eq.~\eqref{function_ML}) is a vector with 2401 elements (401 if we only learn the first 100) and $\boldsymbol{\overline{\overline f}}$ is
\begin{widetext}
\setcounter{MaxMatrixCols}{17}
  \begin{gather}\label{P_mat}
  \boldsymbol{\overline{\overline f}} =
    \setlength{\arraycolsep}{.75\arraycolsep}
    \text{\footnotesize$\displaystyle
    \begin{pmatrix}
                   \left(a_0^>\right)_1 & \ldots & \left(a_{N_c}^>\right)_1 & \left(b_0^>\right)_1 & \ldots & \left(b_{N_c}^>\right)_1 & \left(a_0^<\right)_1 & \ldots & \left(a_{N_c}^<\right)_1 & \left(b_0^<\right)_1 & \ldots & \left(b_{N_c}^<\right)_1 & \left(E_{GS}\right)_1\\
                   \left(a_0^>\right)_2 & \ldots & \left(a_{N_c}^>\right)_2 & \left(b_0^>\right)_2 & \ldots & \left(b_{N_c}^>\right)_2 & \left(a_0^<\right)_2 & \ldots & \left(a_{N_c}^<\right)_2 & \left(b_0^<\right)_2 & \ldots & \left(b_{N_c}^<\right)_2 & \left(E_{GS}\right)_2\\
                   \vdots & \ddots & \vdots & \vdots & \ddots & \vdots & \vdots & \ddots & \vdots & \vdots & \ddots & \vdots & \vdots \\
                   \vdots & \ddots & \vdots & \vdots & \ddots & \vdots & \vdots & \ddots & \vdots & \vdots & \ddots & \vdots & \vdots \\
                   \left(a_0^>\right)_{N_{L}} & \ldots & \left(a_{N_c}^>\right)_{N_{L}} & \left(b_0^>\right)_{N_{L}} & \ldots & \left(b_{N_c}^>\right)_{N_{L}} & \left(a_0^<\right)_{N_{L}} &  \ldots & \left(a_{N_c}^<\right)_{N_{L}} & \left(b_0^<\right)_{N_{L}} & \ldots & \left(b_{N_c}^<\right)_{N_{L}} & \left(E_{GS}\right)_{N_{L}}\\
    \end{pmatrix}.
  $}
  \end{gather}
  \end{widetext}
It is important to note that although the continued fraction representation is a formal way to write any function with a spectral representation, in practice, an accurate numerical process for obtaining the coefficients is available only in the context of ED calculations.
\subsubsection{Matsubara frequency representation}\label{QI_sec_Gwn}
We may evaluate the calculated $G$ on the Matsubara points $\omega_n=(2n+1)\pi T$ (many QMC codes also give $G$ evaluated on these points).  Then
\setcounter{MaxMatrixCols}{17}
  \begin{gather}\label{P_mat_Gwn}
  \boldsymbol{\overline{\overline f}} =
    \setlength{\arraycolsep}{.75\arraycolsep}
    \text{\footnotesize$\displaystyle
    \begin{pmatrix}
                   \left(G(i\omega_0)\right)_1 & \left(G(i\omega_1)\right)_1 & \ldots & \left(G(i\omega_{N_c})\right)_1\\
                   \left(G(i\omega_0)\right)_2 & \left(G(i\omega_1)\right)_2 & \ldots & \left(G(i\omega_{N_c})\right)_2\\
                   \vdots & \vdots & \ddots  & \vdots  \\
                   \left(G(i\omega_0)\right)_{N_L} & \left(G(i\omega_1)\right)_{N_L} & \ldots & \left(G(i\omega_{N_c})\right)_{N_L}\\
    \end{pmatrix}.
  $}
  \end{gather}
\subsubsection{Imaginary time representation}\label{QI_sec_Gtau}
By evaluating Eq.~\eqref{G_cont_frac} on the Matsubara points and then Fourier transforming we obtain the Green's function in imaginary time (this is also a standard output of QMC calculations). $G(\tau)$ is real and smooth. We then approximate the continuous $G(\tau)$ by its values on the $N_c$ discrete points $\tau_j=\beta j/N_c$ with $j=0,\ldots,N_c-1$. The value at $\tau = \beta^-$ does not have to be learned since it can be obtained from Eq.~\eqref{anti_tau}). It is also useful to learn a $N_c+1th$ point, namely, the first derivative at $\tau = 0^+$.  Knowledge of this value helps in evaluation of the reverse Fourier transform\cite{Arsenault_IPTD}. In this work we use $N_c = 2^{11}$ and we write
\begin{widetext}
\setcounter{MaxMatrixCols}{17}
  \begin{gather}\label{P_mat_Gtau}
  \boldsymbol{\overline{\overline f}} =
    \setlength{\arraycolsep}{.75\arraycolsep}
    \text{\footnotesize$\displaystyle
    \begin{pmatrix}
                   \left(G(\tau = 0^+)\right)_1 & \ldots & \left(G(\tau = \beta/2)\right)_1 & \ldots & \left(G(\tau = \beta (1-1/N_{\tau}))\right)_1 & \left(G'(\tau = 0^+)\right)_1\\
                   \left(G(\tau = 0^+)\right)_2 & \ldots & \left(G(\tau = \beta/2)\right)_2 & \ldots & \left(G(\tau = \beta (1-1/N_{\tau}))\right)_2 & \left(G'(\tau = 0^+)\right)_2\\
                   \vdots & \ddots & \vdots & \ddots & \vdots & \vdots  \\
                   \vdots & \ddots & \vdots & \ddots & \vdots & \vdots  \\
                   \left(G(\tau = 0^+)\right)_{N_L} & \ldots & \left(G(\tau = \beta/2)\right)_{N_L} & \ldots & \left(G(\tau = \beta (1-1/N_{\tau}))\right)_{N_L} & \left(G'(\tau = 0^+)\right)_{N_L}\\
    \end{pmatrix}.
  $}
  \end{gather}
  \end{widetext}
\subsubsection{Legendre orthogonal polynomial representation of the Green's function}\label{QI_sec_Gl}
Recently, Boehnke \emph{et al}.\cite{Boehnke_Legendre} proposed to represent the imaginary time Green's function as a sum  of  Legendre polynomials and measure the  coefficients  in a QMC calculation. Legendre polynomials are chosen instead of other sets of orthogonal polynomials because of the simplicity they offer for transformation to the Matsubara axis. Such an expansion acts as a physically motivated low-pass filter that eliminates the large statistical noise at high frequency coming from the direct calculation of $G$ in Matsubara frequency well known in continuous time QMC (CTQMC). The standard Legendre polynomials $P_l(x)$ are defined on an interval $x$ $\epsilon$ $[-1,1]$. In the case of the Green's function in positive imaginary time, the time interval is $0 < \tau < \beta$ so that we may define the variable $x=\d{2\tau}{\beta}-1$. Thus (see Appendix~\ref{Legendre_appen})
\begin{equation}\label{Leg_exp_Gtau}
  G(\tau) = \sum_{l=0}^{\infty}\d{\sqrt{2l+1}}{\beta}G_lP_l(x(\tau)),
\end{equation}
where the coefficients are formally given by
\begin{equation}\label{coeff_Leg_Gtau}
  G_l = \sqrt{2l+1}\int_0^{\beta}d\tau P_l(x(\tau))G(\tau).
\end{equation}
The Fourier transform to $\omega_n$ is  given by [\onlinecite{Boehnke_Legendre}] as
\begin{equation}\label{Leg_exp_Giwn}
  G(i\omega_n) = \sum_{l=0}^{\infty} T_{nl}G_l,
\end{equation}
where
\begin{equation}\label{T_coeff}
  T_{nl} = (-1)^{n}i^{l+1}\sqrt{2l+1}j_l\left(\d{(2n+1)\pi}{2}\right),
\end{equation}
and $j_l(z)$ are the spherical Bessel functions. $G_l$ may be directly  measured in a CTQMC calculation \cite{Boehnke_Legendre} and the maximum order $l_{max}$ is defined as the largest $l$ where $G_l$ is greater than the  statistical noise.  In the present case we must calculate the $G_l$ and devise an alternative prescription for $l_{max}$. We use a recently introduced algorithm based on a fast Chebyshev-Legendre transform\cite{Hale_2014} that exploits the idea that smooth functions can be represented by polynomial interpolation in Chebyshev points, i.e., by expansions in Chebyshev polynomials using fast fourier transform.  This algorithm is implemented in a free Matlab toolbox called CHEBFUN\cite{chebfun}. We still have to define $l_{max}$. The $l_{max}$ is chosen by looking at the odd Legendre coefficients. For every example we have investigated, we find that  $G_l$ for $l$ odd decreases rapidly as $l$ increases. At some $l$, $G_l$ changes sign and starts oscillating around zero.  We set $l_{max}$ by finding the $l$ at which  $G_l$ changes sign. For each example (a particular set of parameters $U$, $V$, and $Wn_d$), the value of $l$ for which the sign change happens is different. However, for the 5000 examples in the database, we found that the first sign change happens for $l$ at most around 111. For security, we use as our definition of the expansion the first 121 ($l = 0\ldots 120$) terms. For each example, for $l \leq l_{max}$ for the odd $l$ we use the values we have while we have to decide what values to use for $l > l_{max}$ (for $l$ even we do not change anything). Two easily implemented options are to replace the odd coefficients by zero for $l > l_{max}$ or to replace them by the last value before the first sign change of $G_l$ so that either $G_l = 0$ for $l$ odd $> l_{max}$ or $G_l = cst$. We verified that both solutions work very well to reconstruct $G(\tau)$ from Eq.~\eqref{Leg_exp_Gtau}. However, for machine learning there is a difference. Indeed, in ML, having pure zeros in the learning set makes the learning process much more difficult. We thus use the second solution where $G_l$ for $l$ odd $> l_{max}$ is a very small constant.\\
\\
The great advantage of the Legendre polynomial representation is that either by obtaining $G_l$ from CTQMC or directly from $G(\tau)$, the number of coefficients is very limited, making the learning process much smaller, and, if after learning all coefficients at once, we see that some fine tuning is needed, the maximum number of new machines that will be necessary is limited and manageable. Also, as we do not directly learn the function $G(\tau) (G(i\omega_n))$ but reconstruct it from Eq.~\eqref{Leg_exp_Gtau} (or Eq.~\eqref{Leg_exp_Giwn}) it helps smooth things out.\\
\\
Therefore what is directly learned is the vector of coefficients.
\setcounter{MaxMatrixCols}{17}
  \begin{gather}\label{P_mat_Gl}
  \boldsymbol{\overline{\overline f}} =
    \setlength{\arraycolsep}{.75\arraycolsep}
    \text{\footnotesize$\displaystyle
    \begin{pmatrix}
                   \left(G_0\right)_1 & \left(G_1\right)_1 & \ldots & \left(G_{120}\right)_1\\
                   \left(G_0\right)_2 & \left(G_1\right)_2 & \ldots & \left(G_{120}\right)_2\\
                   \vdots & \vdots & \ddots & \vdots \\
                   \left(G_0\right)_{N_L} & \left(G_1\right)_{N_L} & \ldots & \left(G_{120}\right)_{N_L}\\
    \end{pmatrix}.
  $}
  \end{gather}
\section{Results}\label{results}
In this section we present the results both of computation of the full $G$ and, for two particular examples,  of the renormalization factor $Z$. For the two examples we will use parameters similar to those used in [\onlinecite{Krivenko}] and [\onlinecite{Kirchner}]. In one case [\onlinecite{Krivenko}] $U = 3$, $V = 0.5$, and $\epsilon_d = -U/2$. In the second case [\onlinecite{Kirchner}], an infinite $U$ AIM is considered with a flat conduction-band DOS with $V = \sqrt{0.1}$ and $\epsilon_d = -0.81$. In the case of infinite $U$ [\onlinecite{Kirchner}], this gives an occupation $n_d \approx 0.94$. We will use the two examples in our database that are the closest to these parameters. For the first case, we have $U = 3.04$, $V = 0.5105$, and $n_d = 1$. For the second case, as infinite $U$, we will use $U = 4$, $V = 0.3053$, and $n_d = 0.95$.  We present the results both in $\omega_n$ and $\omega$. For real frequencies, we need to choose a small imaginary part $\eta$ for the frequency. Since our ED fitting procedure relies on choosing a fictitious temperature to fit on the Matsubara axis defined by this temperature, we take the  minimum-energy unit to be twice the difference between two $\omega_n$ ($= \d{2\pi}{\beta}$). For small dataset length, we generated many random combinations for the learning set, predicted the results for every combination and averaged out.

\subsection{Continued fraction representation}\label{QI_sec_cont_frac_results}
\begin{figure}[tpb]
  \begin{center}
  \includegraphics[scale=0.60]{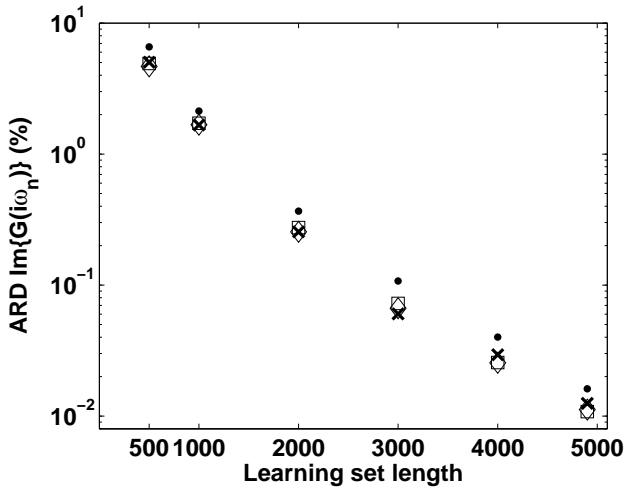}
  \end{center}
  \caption{Average relative difference (ARD) for the predicted imaginary part of the Green's function on the Matsubara axis as a function of training set size. Dots (.) denote the predictions of the continued fraction coefficients, x's (x) denote the predictions of $G(i\omega_n)$, squares ($\square$) denote the predictions of $G(\tau)$, and diamonds ($\diamondsuit$) denote the Legendre polynomial expansion.}
  \label{fig:ARD_coeffs}
\end{figure}

\begin{figure*}[tpbh]
  \begin{center}
  \mbox{\includegraphics[scale=0.5]{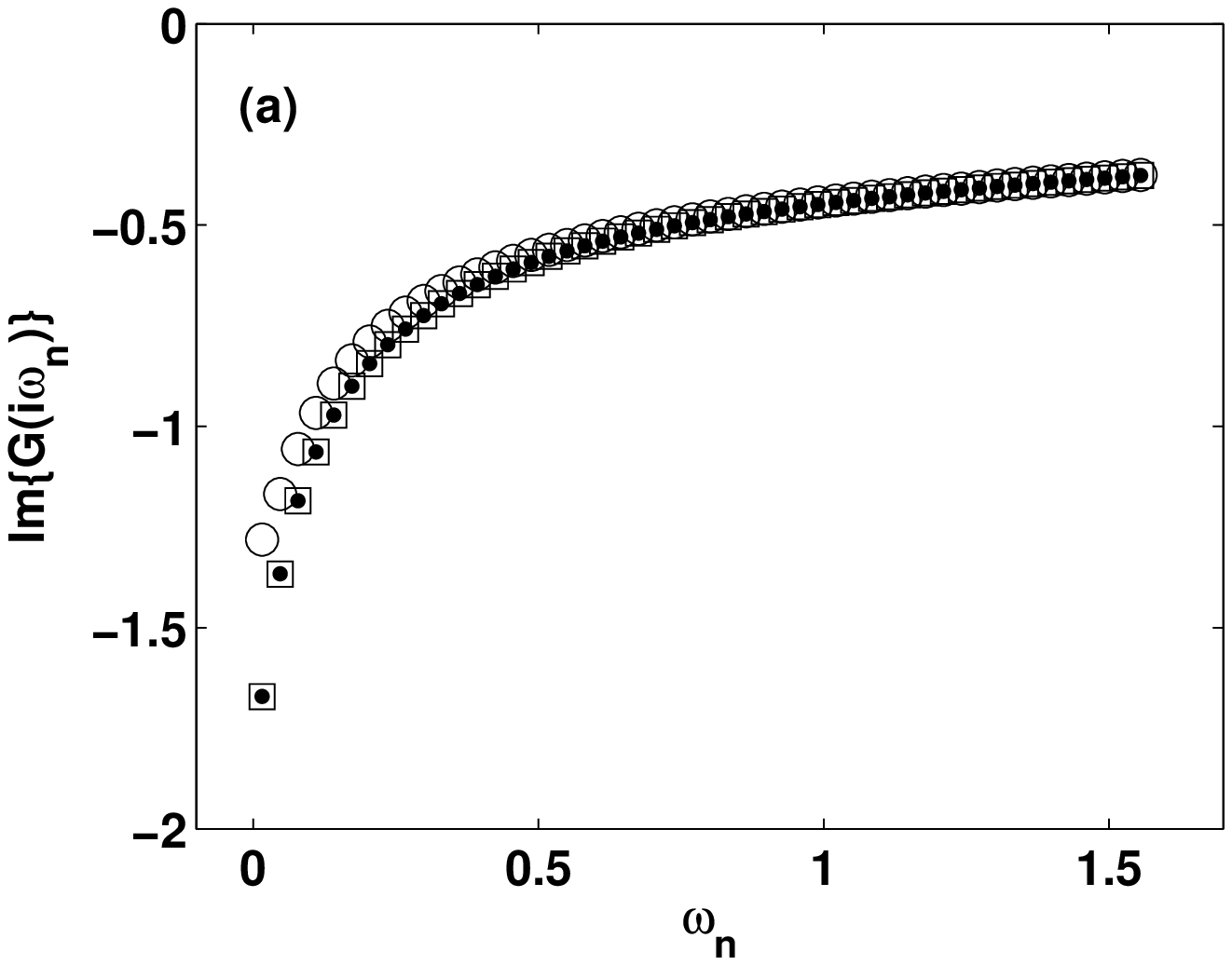}}
  \mbox{\includegraphics[scale=0.5]{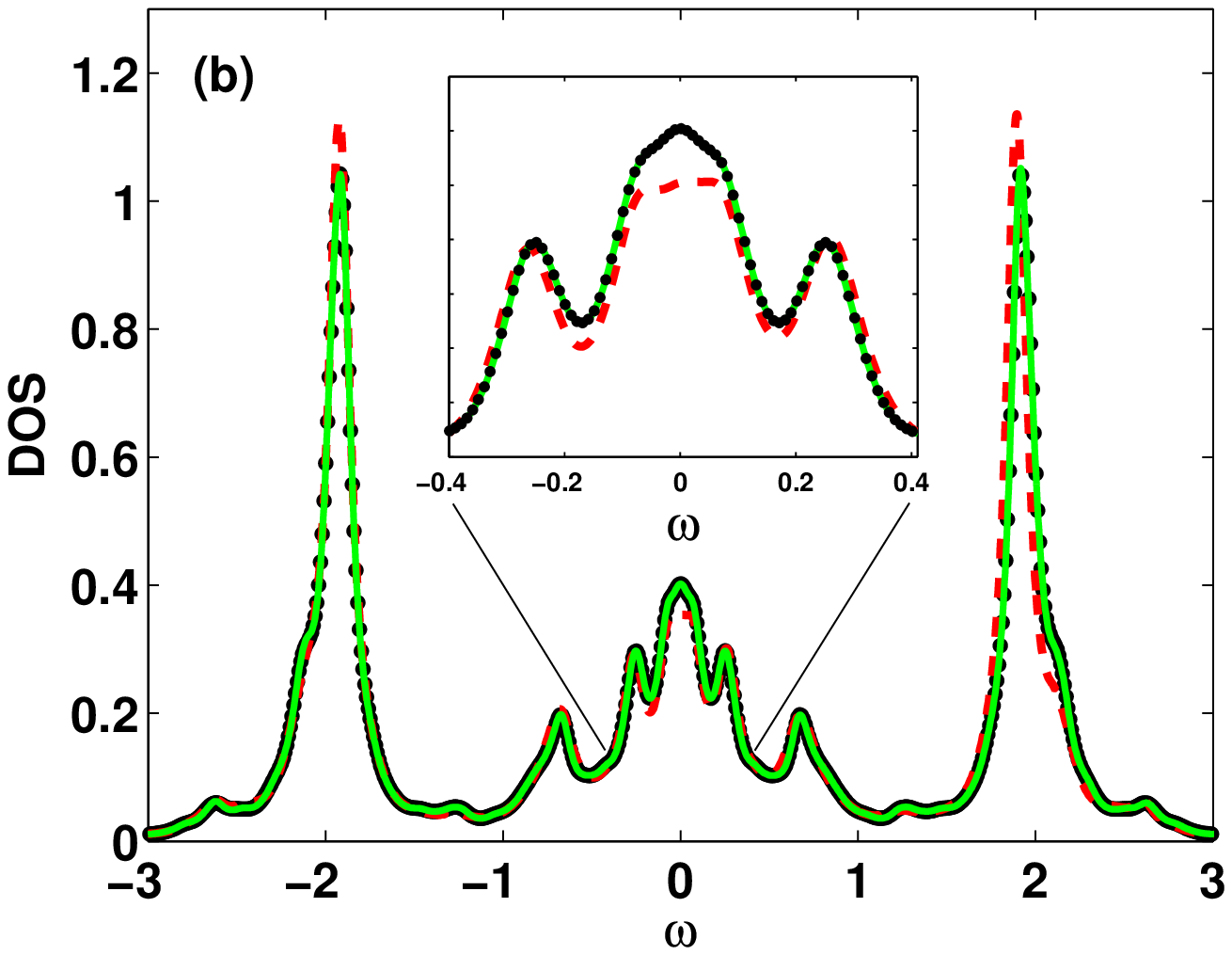}}
  \mbox{\includegraphics[scale=0.5]{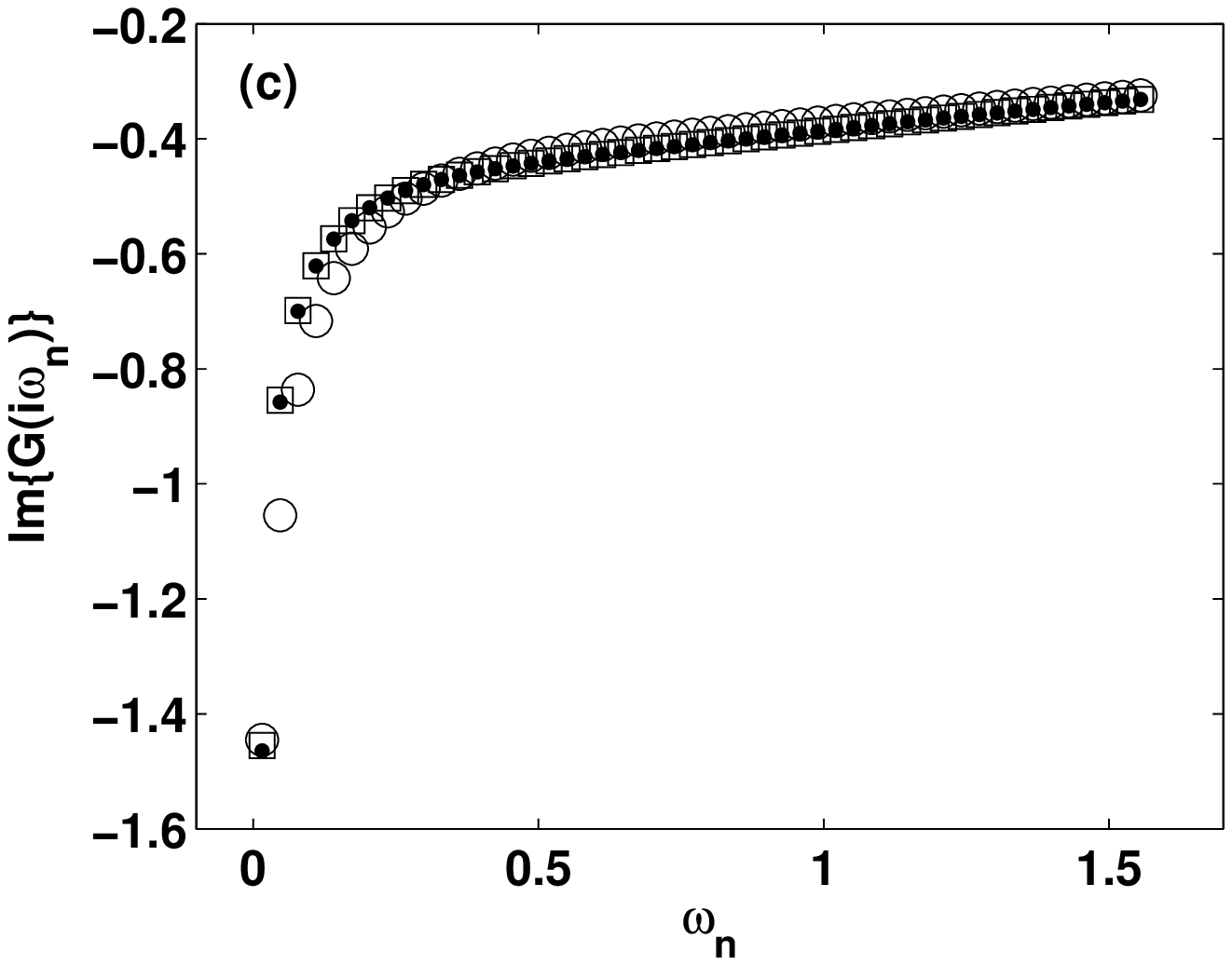}}
  \mbox{\includegraphics[scale=0.5]{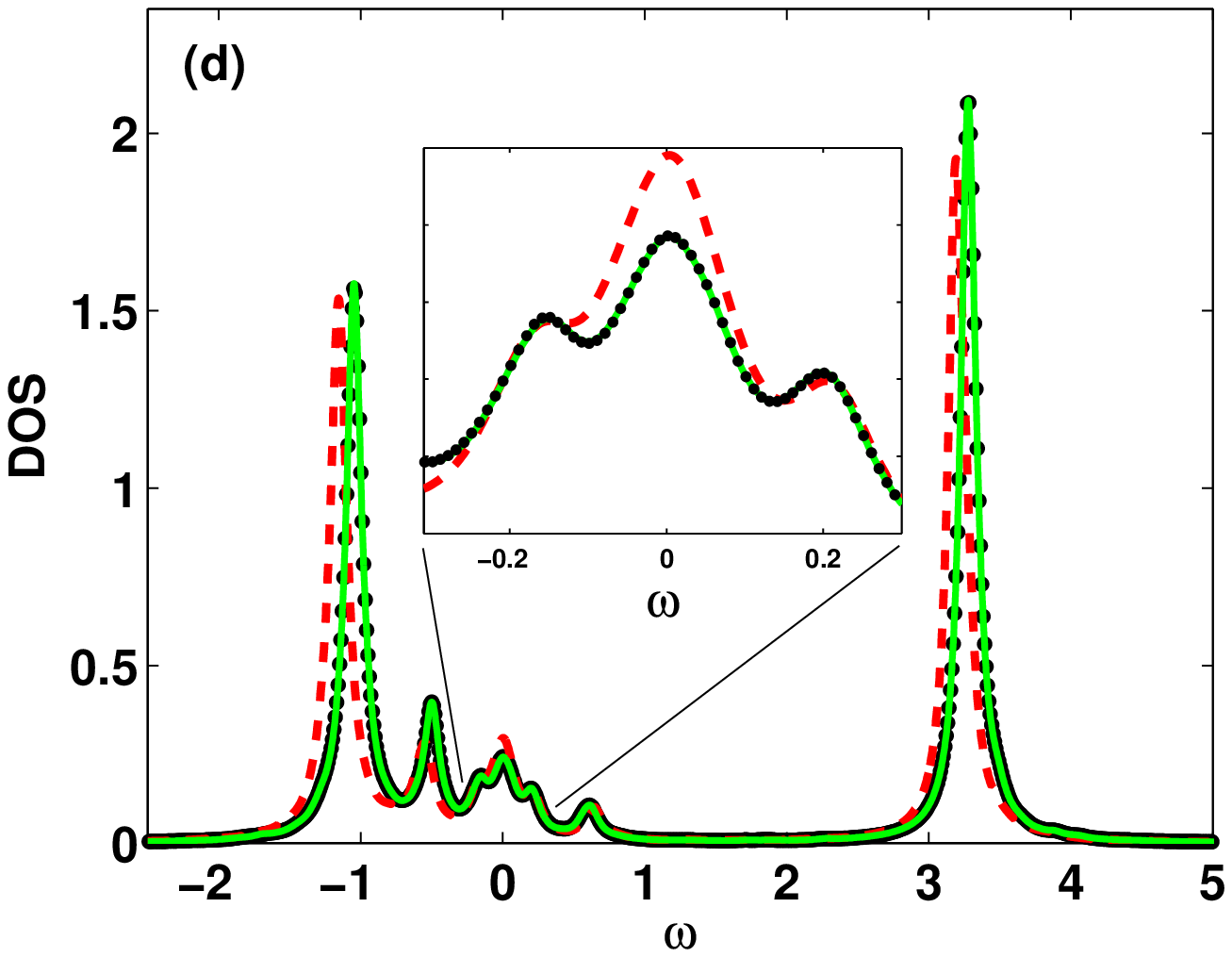}}
  \end{center}
  \caption{(Color online) Machine learning prediction for $\text{Im}\{G(i\omega_n)\}$ and the density of states using the continued fraction coefficients: (a) and (b) for $U = 3.04$, $V = 0.5105$, and $n_d = 1$ and (c) and (d) for $U = 4$, $V = 0.3053$, and $0.95$. In (a) and (c), dots (.) denote the exact result, circle (O) denote the result for a learning set of length 500, and squares ($\square$) denote the result for a learning set of length 4999. In (b) and (d), dots (.) denote the exact result, red dashed lines (- -) denote the result for a learning set of length 500, and green lines (-) denote the result for a learning set of length 4999.}
  \label{fig:predic_wn_w_coeffs}
\end{figure*}

We first use ML to learn the coefficients of the continued fraction representation of the Green's function (Eq.~\eqref{G_cont_frac}). We learned the first hundred of each type since, as we mentioned in Sec.~\ref{QI_AIM}, the remaining coefficients do not contribute to $G$ within our accuracy.  Using test sets of length 100 and generating learning and test sets multiple times, from the reconstructed $G$ we calculate the ARD, and we show the results as a function of training set size in Fig.~\ref{fig:ARD_coeffs} as dots. We see that the ARD decreases from about 6.5\% for a random learning set of 500 examples to about 0.016\% for a learning set of 4900.\\
\\
For the comparison with the two specific examples, the results are presented in Fig.~\ref{fig:predic_wn_w_coeffs}. Only the first 50 Matsubara frequencies are shown for $\text{Im}\{G(i\omega_n)\}$.\\
\\
For the half-filled case, Fig.~\ref{fig:predic_wn_w_coeffs}-(a) and (b), we see that the prediction for randomly chosen learning sets of 500 in $\omega_n$ is not very good. In terms of real frequency, we see that, however, even with the size 500 random learning set, the high-frequency regions are well predicted. However, around the Fermi level, the prediction is wrong. For example, the prediction is not particle-hole symmetric even though the model is.   However, at the largest learning set size (4999) the prediction is correct. The doped case is similar. From Fig.~\ref{fig:predic_wn_w_coeffs}(c) we see that for this particular example, the value at the lowest Matsubara frequency is well predicted for a small random learning set but the values at the next three frequencies are not good. Turning now to real frequencies we see that the high-frequency regions and the region around the Fermi level are qualitatively fairly predicted. For a large learning set, the doped case is also well represented.\\
\\
Thus, if the Green's function is learned from its continued fraction coefficients, a small and random learning set is able to capture the high-frequency physics for the half-filled case, but it is not reliable for the low $\omega$ which carry most of the interesting physical information at $T = 0$.
\subsection{Matsubara frequency representation}\label{QI_sec_Gwn_results}
\begin{figure*}[tpbh]
  \begin{center}
  \mbox{\includegraphics[scale=0.5]{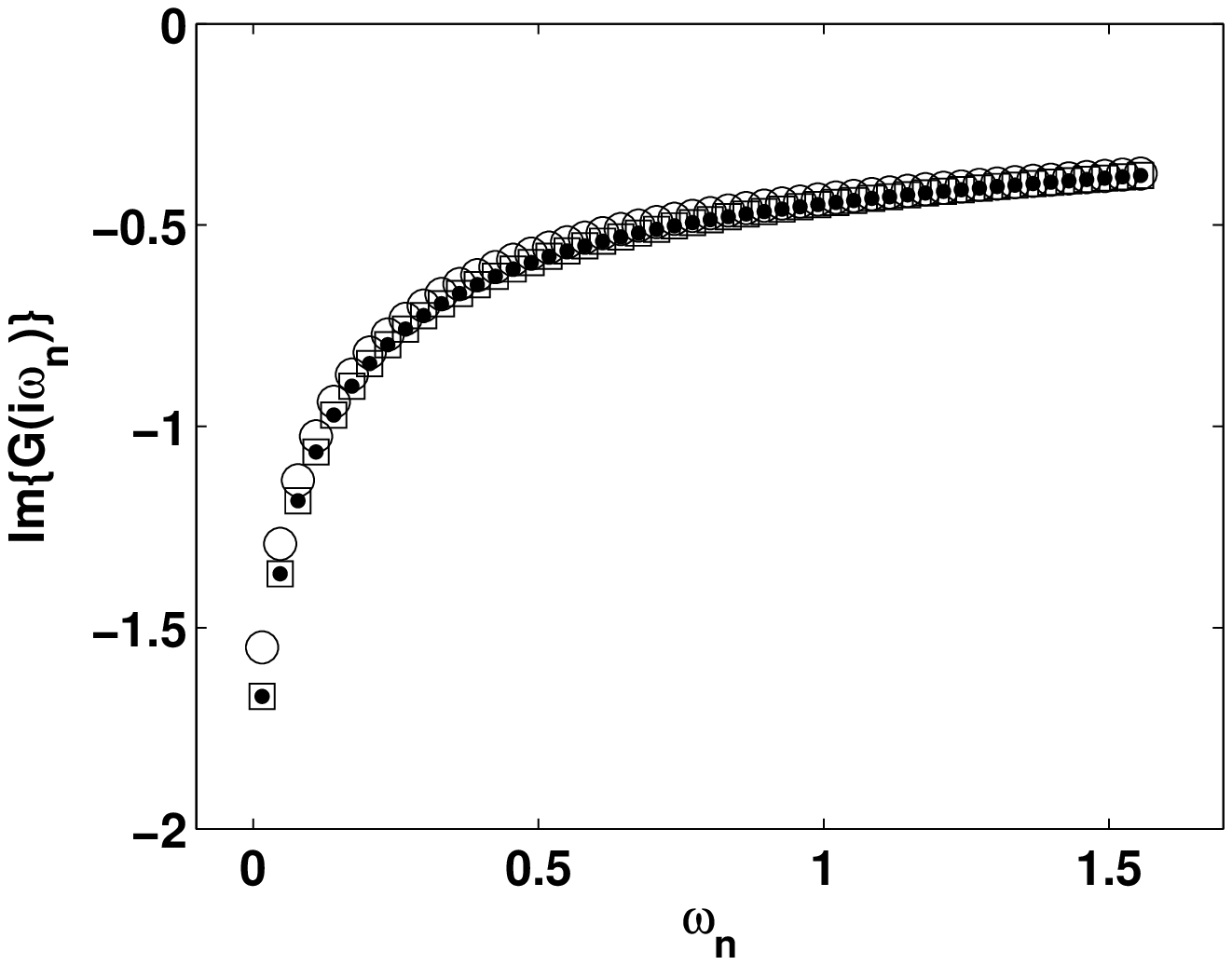}}
  \mbox{\includegraphics[scale=0.5]{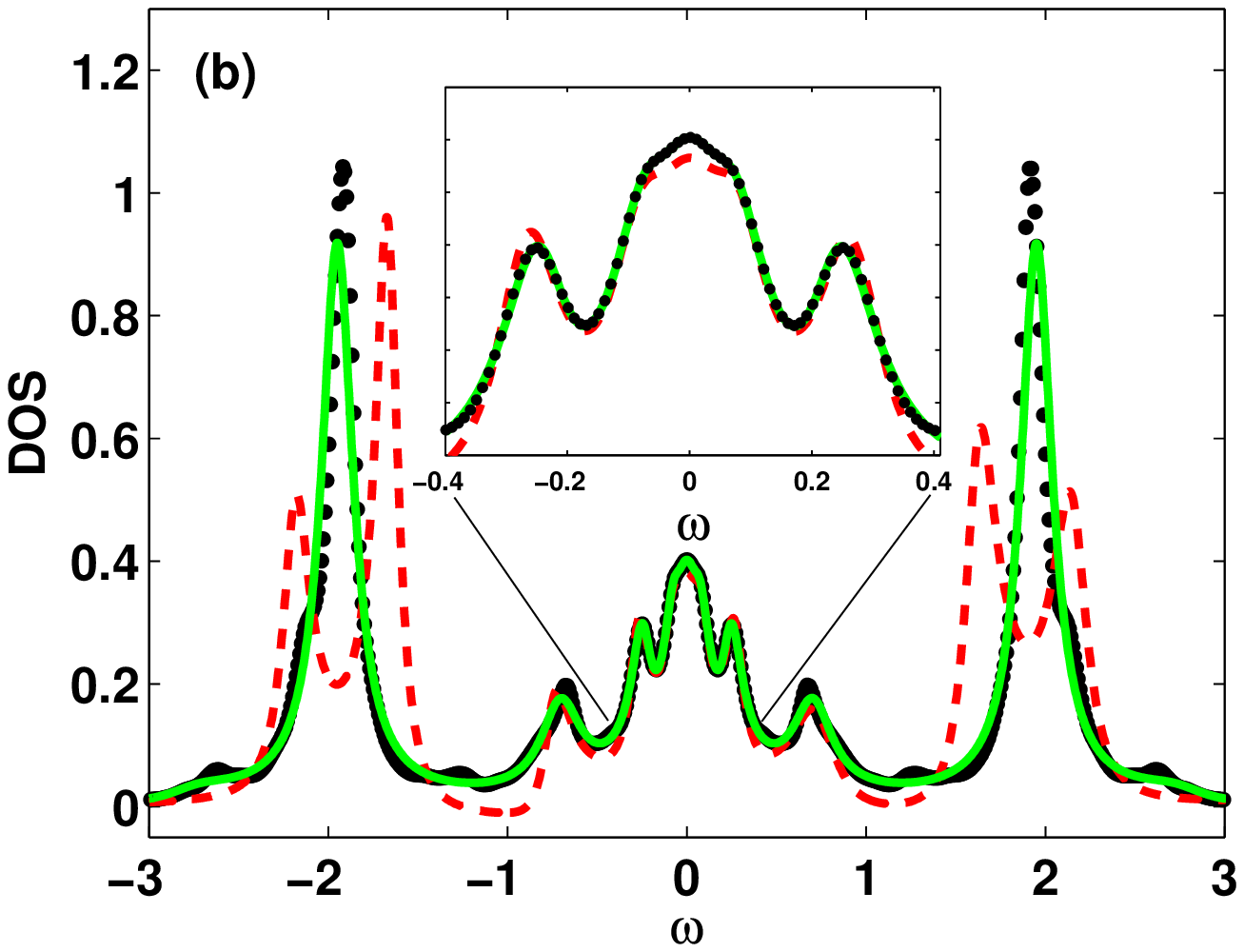}}
  \mbox{\includegraphics[scale=0.5]{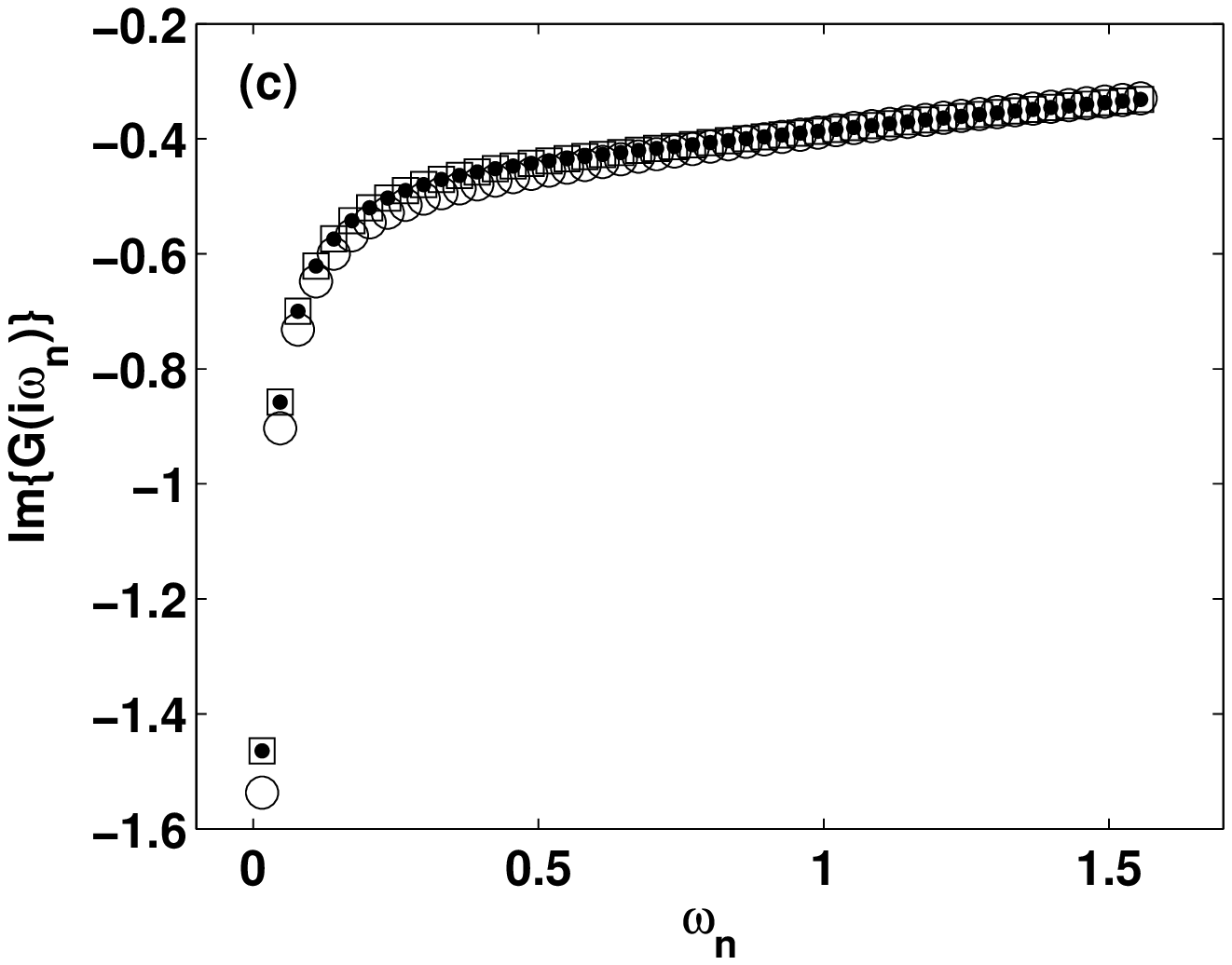}}
  \mbox{\includegraphics[scale=0.5]{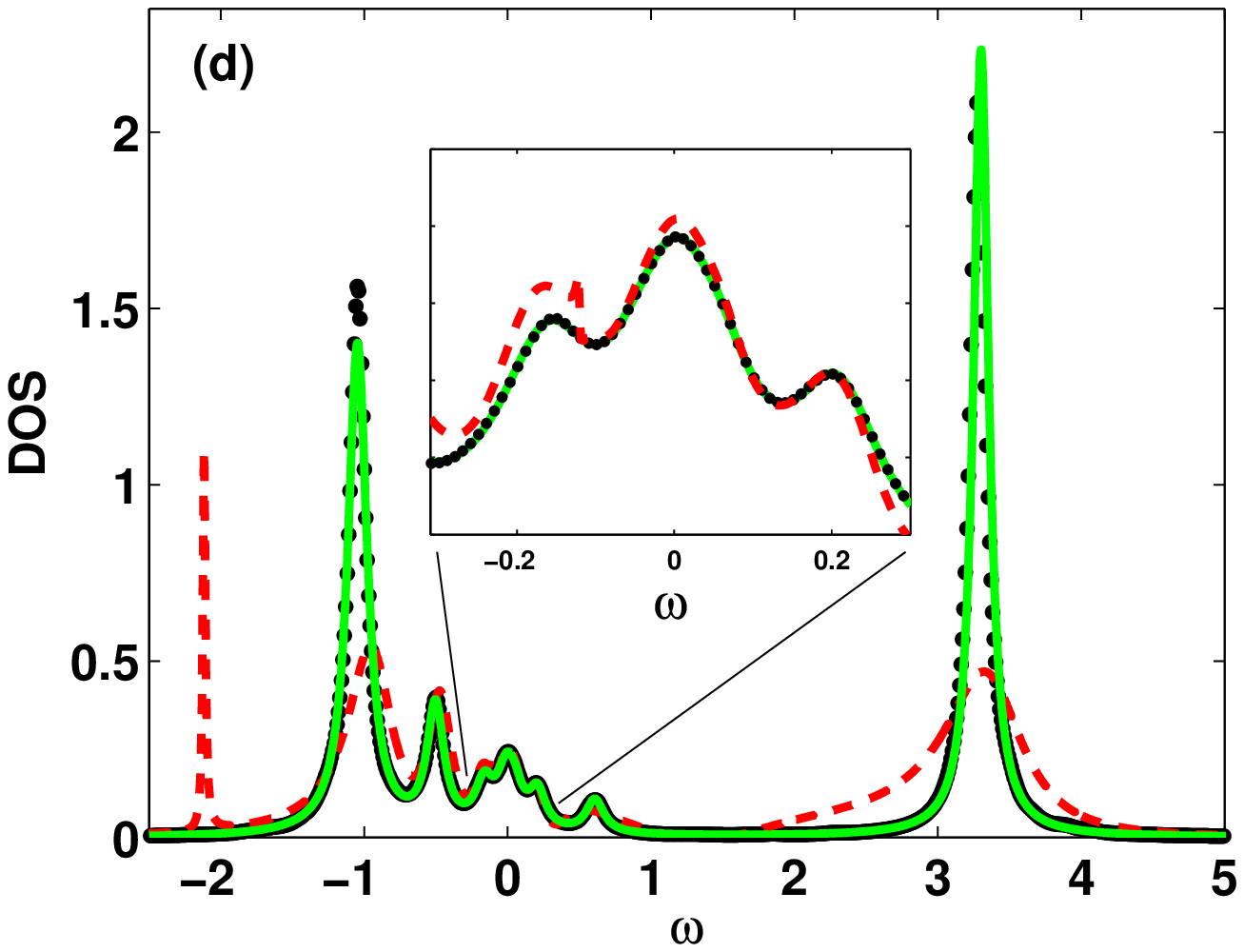}}
  \end{center}
  \caption{(Color online) machine learning prediction for $\text{Im}\{G(i\omega_n)\}$ and the density of states using $G(i\omega_n)$ directly: (a) and (b) for $U = 3.04$, $V = 0.5105$, and $n_d = 1$, and (c) and (d) for $U = 4$, $V = 0.3053$, and $0.95$. In (a) and (c), dots (.) denote the exact result, circles (O) denote the result for a learning set of length 500, and squares ($\square$) denote the result for a learning set of length 4999. In (b) and (d), dots (.) denotes the exact result, red dashed lines (- -) denote the result for a learning set of length 500, and green lines (-) denote the result for a learning set of length 4999.}
  \label{fig:predic_wn_w_Gwn}
\end{figure*}

The ARDs are denoted by x's in Fig.~\ref{fig:ARD_coeffs}. The values are systematically  smaller than the equivalent from the continued fraction coefficients, but still the same order of magnitude. The real frequency result at half filling of Fig.~\ref{fig:predic_wn_w_Gwn}(b) shows that this time, around the Fermi level, the ML predicted DOS has the correct qualitative behavior of particle-hole symmetry. If we compare the Matsubara frequency results of Fig.~\ref{fig:predic_wn_w_Gwn}(a) and (c) with those of Fig.~\ref{fig:predic_wn_w_coeffs}(a) and (c) we see that for these two examples, learning directly $G(i\omega_n)$ is better. However, the analytically continued results for the learning set of length 500 are of poor quality, essentially because analytical continuation is sensitive to small errors in $G(i\omega_n)$. This is thus hard to assess if the prediction really contains noncausality (Fig.~\ref{fig:predic_wn_w_Gwn}(b)) ($ -1.5 < \omega < -1$), gives such shape at high frequency (Fig.~\ref{fig:predic_wn_w_Gwn}(b) and (d)) and give spurious states where the DOS should be zero (Fig.~\ref{fig:predic_wn_w_Gwn}(d) at about $\omega = -2.1$). Once again, for a large learning set, the predicted results are very good. It is important to note that since we need numerical analytical continuation to obtain the real frequency results, 100\% perfect matching is impossible.\\
\subsection{Imaginary time representation}\label{QI_sec_Gtau_results}
\begin{figure*}[tpbh]
  \begin{center}
  \mbox{\includegraphics[scale=0.5]{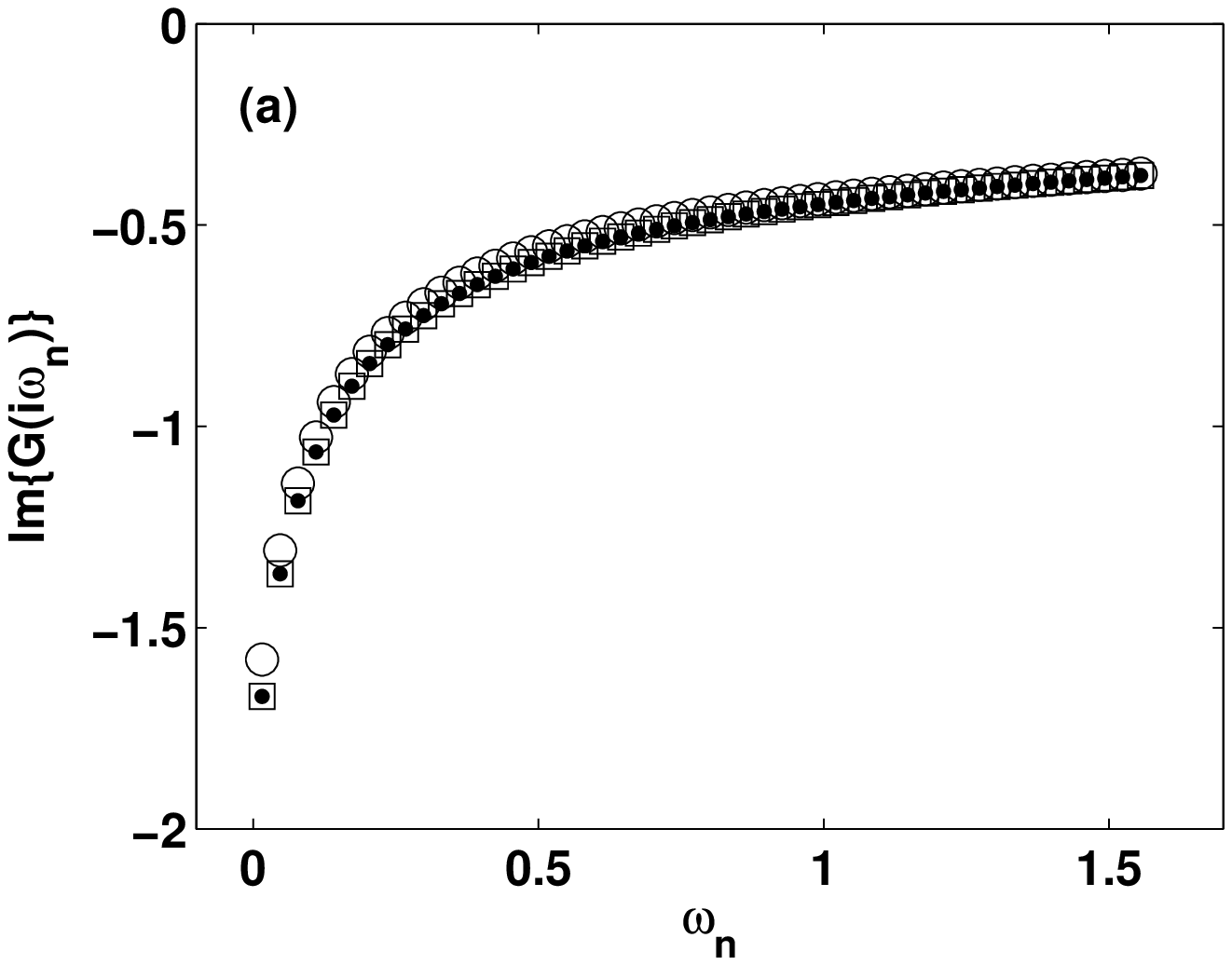}}
  \mbox{\includegraphics[scale=0.5]{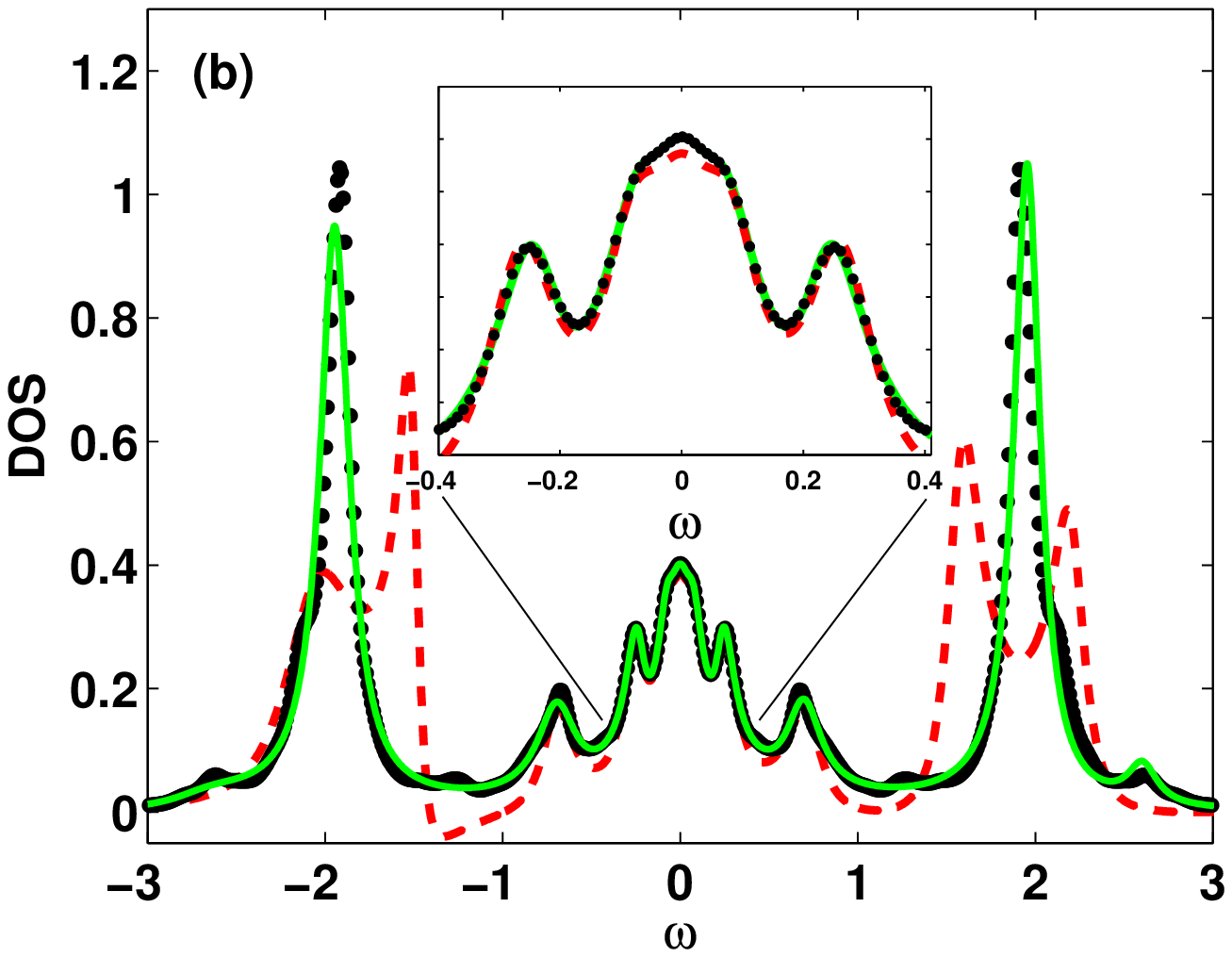}}
  \mbox{\includegraphics[scale=0.5]{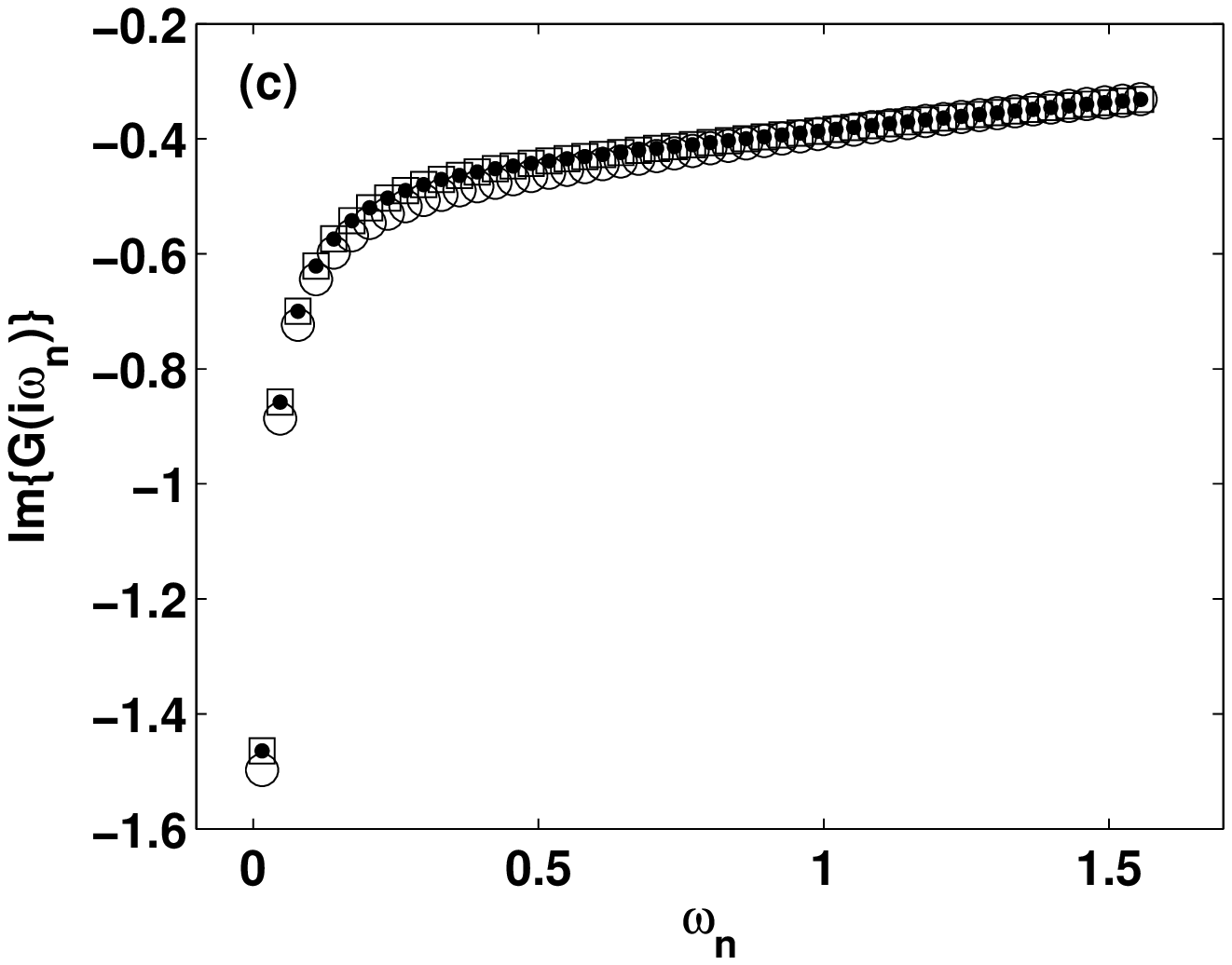}}
  \mbox{\includegraphics[scale=0.5]{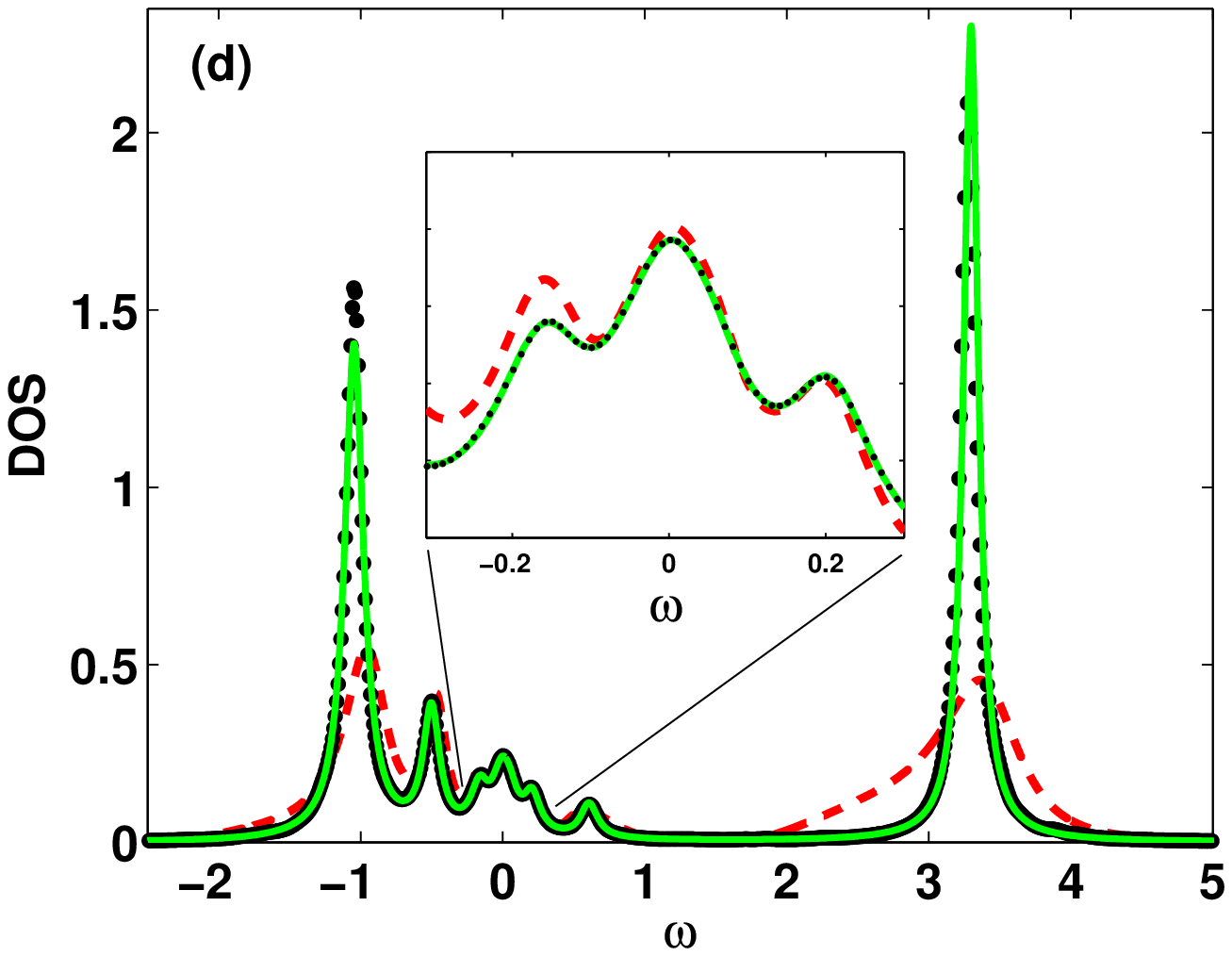}}
  \end{center}
  \caption{(Color online) machine learning prediction for $\text{Im}\{G(i\omega_n)\}$ and the density of states using $G(\tau)$: (a) and (b) for $U = 3.04$, $V = 0.5105$, and $n_d = 1$ and (c) and (d) for $U = 4$, $V = 0.3053$, and $0.95$. In (a) and (c), dots (.) denote the exact result, circles (O) denote the result for a learning set of length 500, and squares ($\square$) denote the result for a learning set of length 4999. In (b) and (d), dots (.) denote the exact result, red dashed lines (- -) denote the result for a learning set of length 500, and green lines (-) denote the result for a learning set of length 4999.}
  \label{fig:predic_wn_w_Gtau}
\end{figure*}

The ARDs for imaginary time representation are denoted by squares in Fig.~\ref{fig:ARD_coeffs}. The values are systematically smaller than the corresponding ARDs from the continued fraction coefficients but still of the same order of magnitude and similar to those from the representation in Matsubara frequency. Most comments made regarding Fig.~\ref{fig:predic_wn_w_Gwn} can also be made for Fig.~\ref{fig:predic_wn_w_Gtau}.
\\
\\
Based on the ARD and the results in $\omega_n$ we can conclude that directly learning $G(\tau)$ is a good procedure and of general applicability.
\subsection{Legendre orthogonal polynomial representation of the Green's function}\label{QI_sec_Gl_results}
\begin{figure*}[tpbh]
  \begin{center}
  \mbox{\includegraphics[scale=0.5]{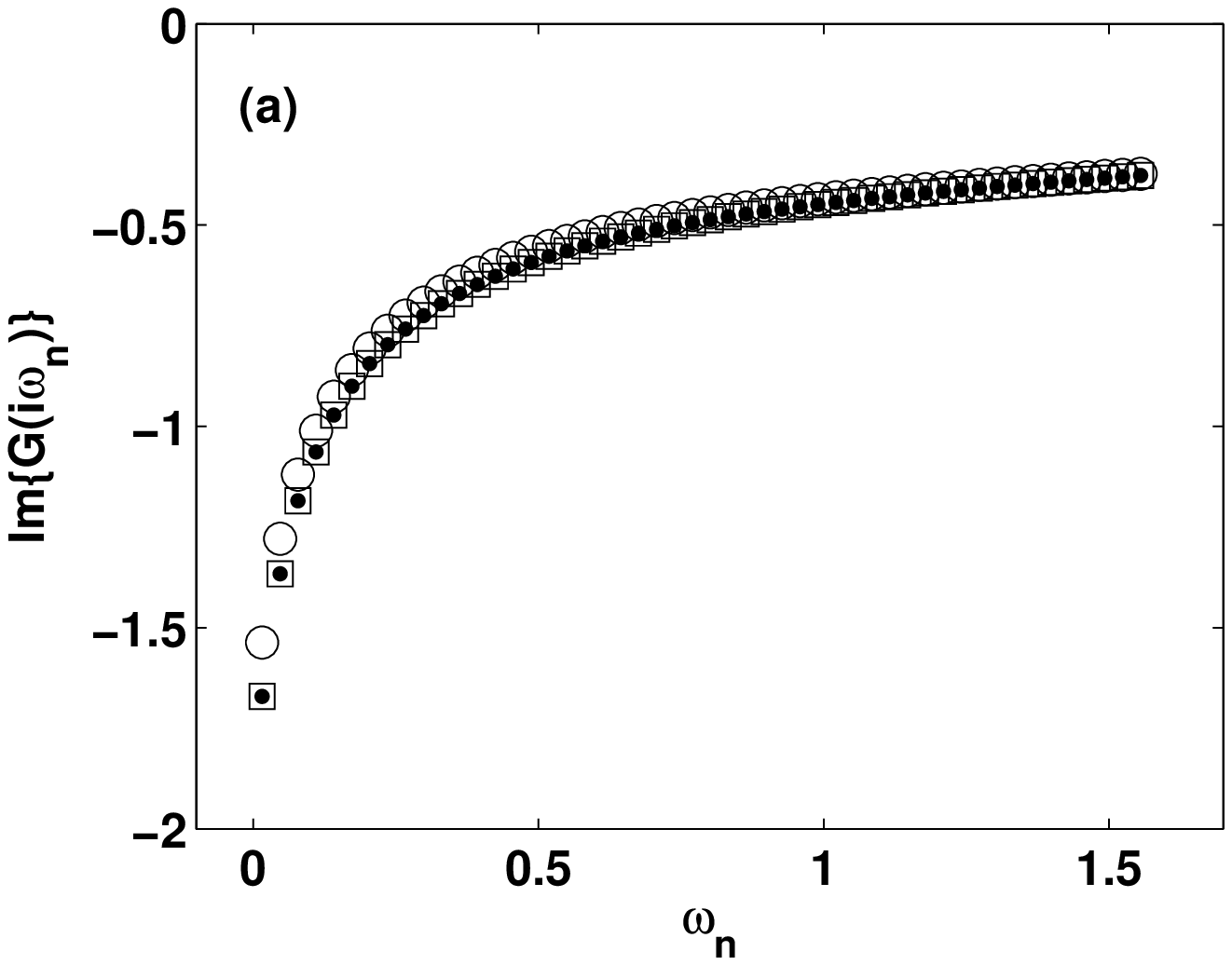}}
  \mbox{\includegraphics[scale=0.5]{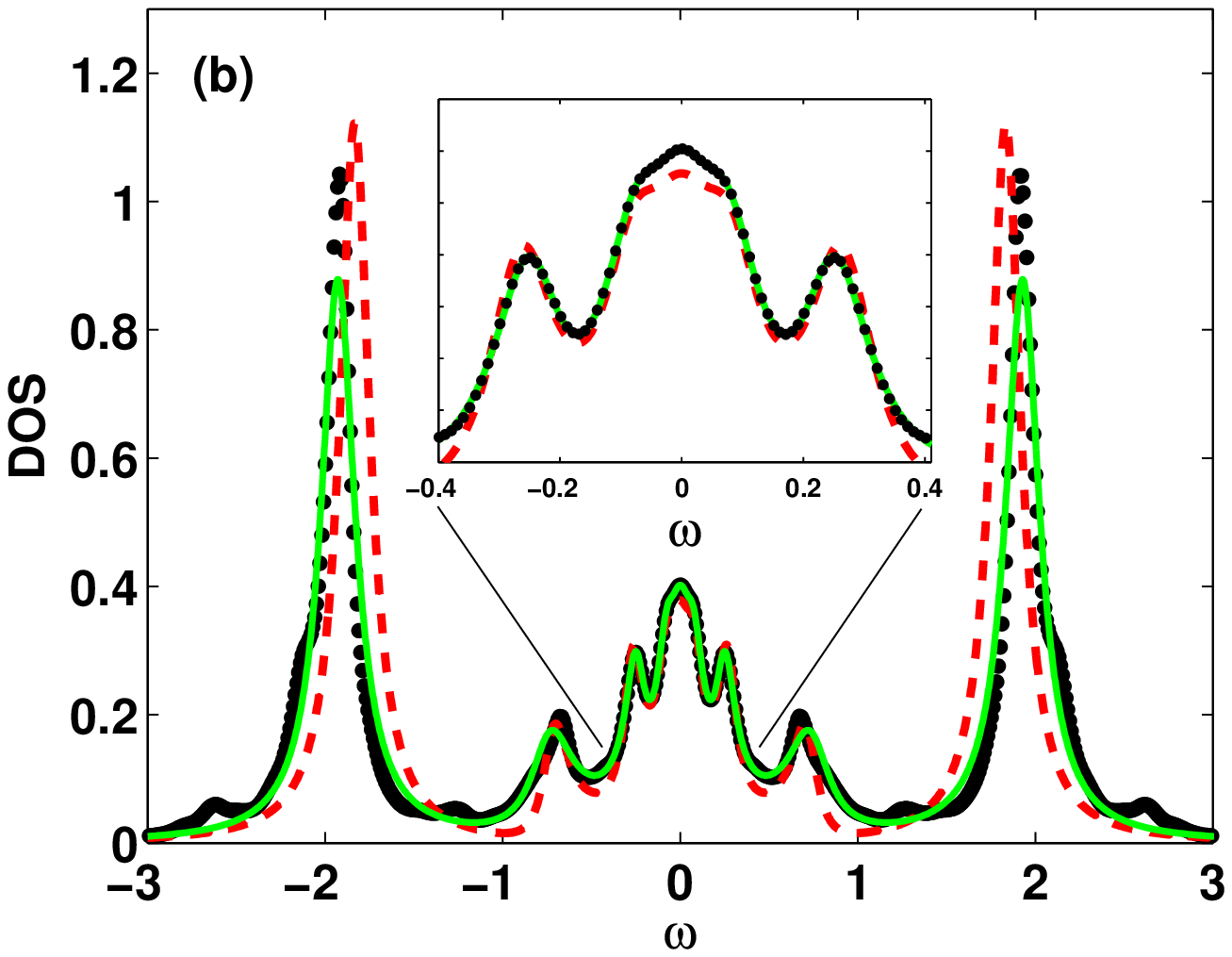}}
  \mbox{\includegraphics[scale=0.5]{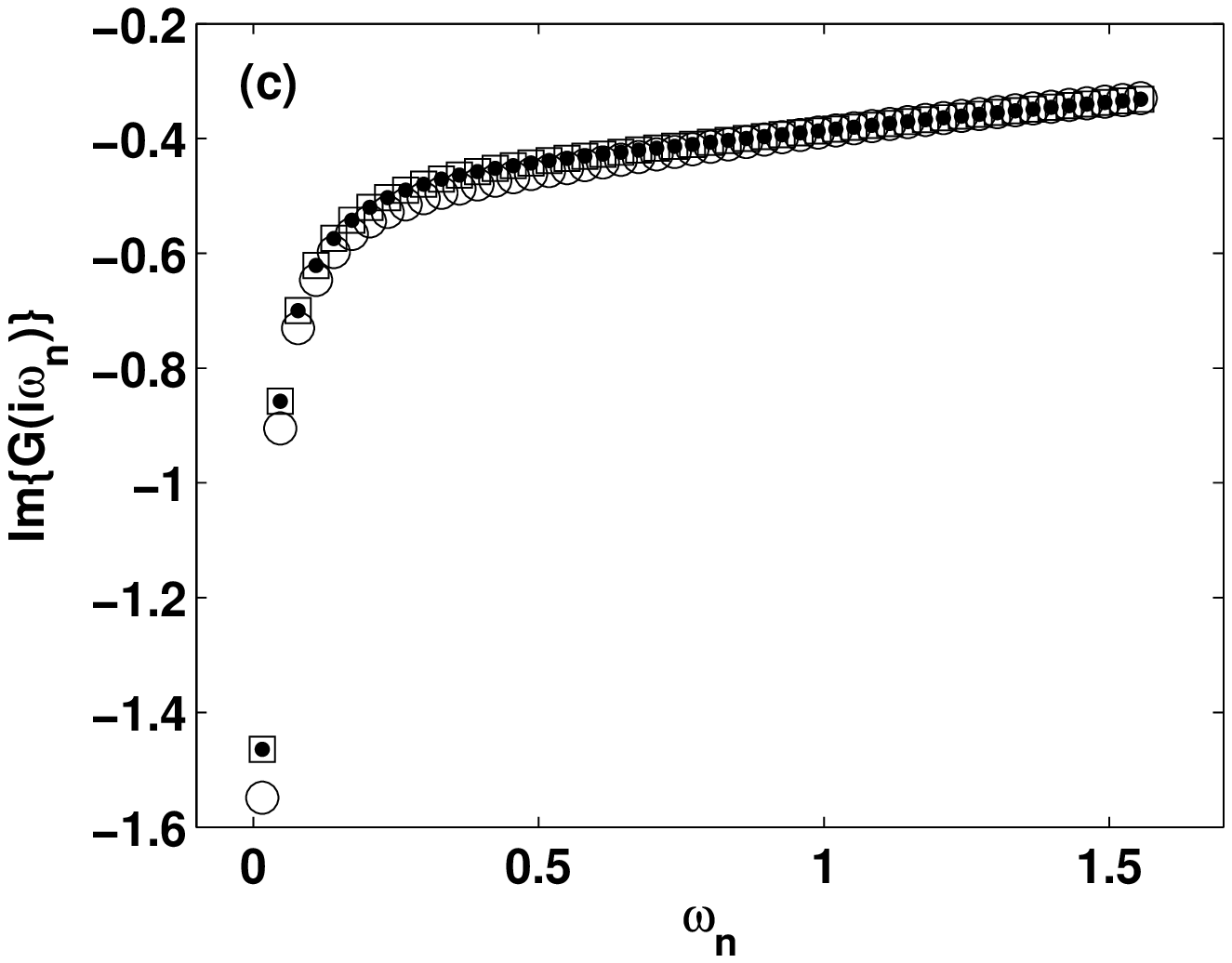}}
  \mbox{\includegraphics[scale=0.5]{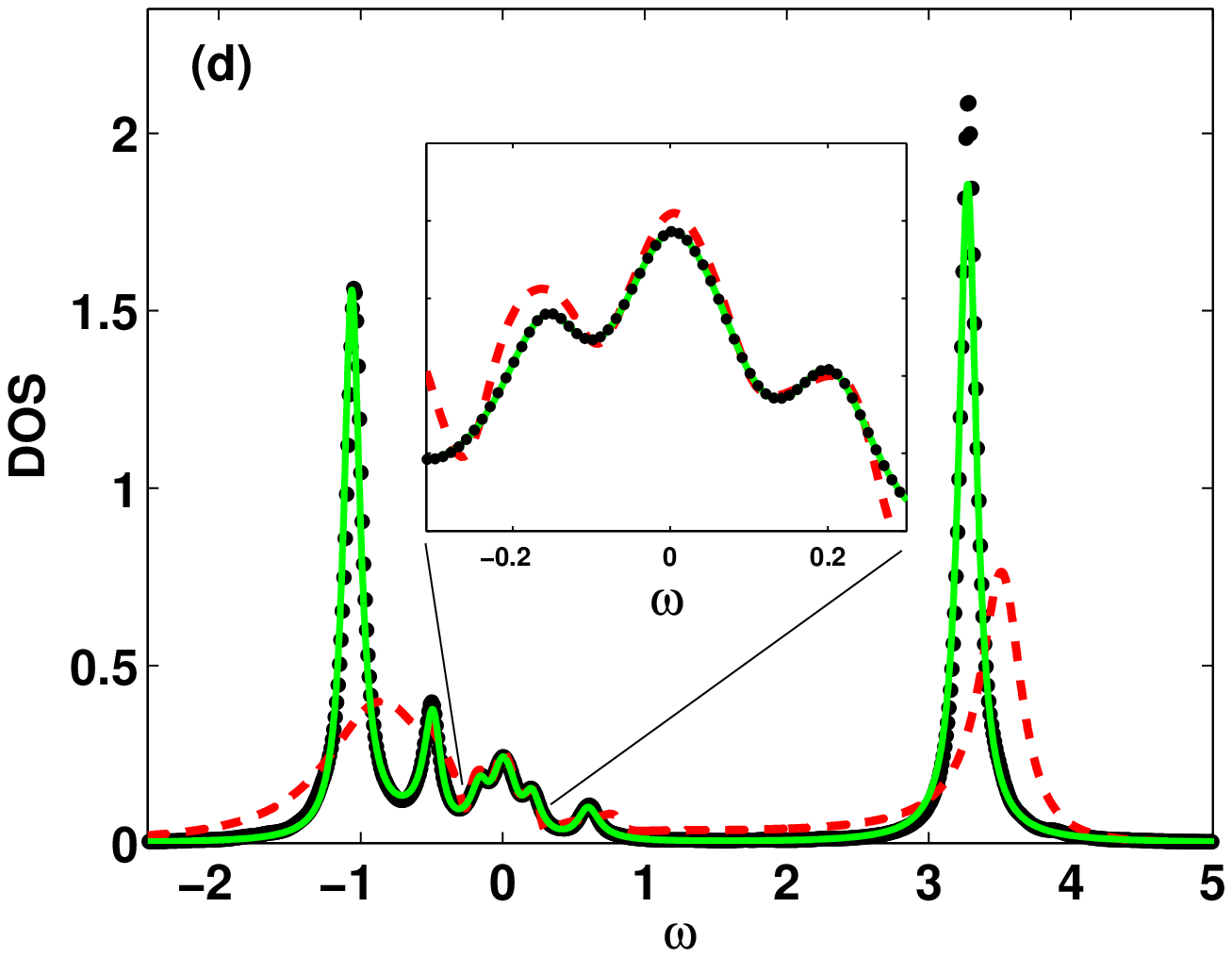}}
  \end{center}
  \caption{(Color online) machine learning prediction for $\text{Im}\{G(i\omega_n)\}$ and the density of states using the Legendre polynomial representation: (a) and (b) for $U = 3.04$, $V = 0.5105$, and $n_d = 1$ and (c) and (d) for $U = 4$, $V = 0.3053$, and $0.95$. In (a) and (c), dots (.) denote the exact result, circles (O) denote the result for a learning set of length 500, and squares ($\square$) denote the the result for a learning set of length 4999. In (b) and (d), dots (.) denote the exact result, red dashed lines (- -) denote the result for a learning set of length 500, and green lines (-) denote the result for a learning set of length 4999.}
  \label{fig:predic_wn_w_Gl}
\end{figure*}

The ARDs of the Legendre orthogonal polynomials representation are denoted by diamonds in Fig.~\ref{fig:ARD_coeffs}. The values are very similar to those from $G(\tau)$ and $G(i\omega_n)$. The results for the two examples in Matsubara and real frequency presented in Fig.~\ref{fig:predic_wn_w_Gl} are comparable to what is obtained from learning directly $G(\tau)$, but this time the Pad\'{e} continuation is less problematic because the coefficients are learned and then the Green's function is reconstructed, giving a $G(i\omega_n)$ with less independent error on each $\omega_n$. Moreover, in Sec.~\ref{QI_sec_Gtau}, we had to learn 2048 slices of $G(\tau)$ plus $G'(\tau = 0^+)$ while here we only needed to learn 121 coefficients.\\
\\
We may therefore conclude that the representation by an expansion in terms of Legendre polynomials is the most efficient way to learn many-body Green's functions using machine learning.
\subsection{Prediction as a function of learning set length}\label{QI_sec_prediction_size}
\begin{figure*}[tpbh]
  \begin{center}
  \mbox{\includegraphics[scale=0.5]{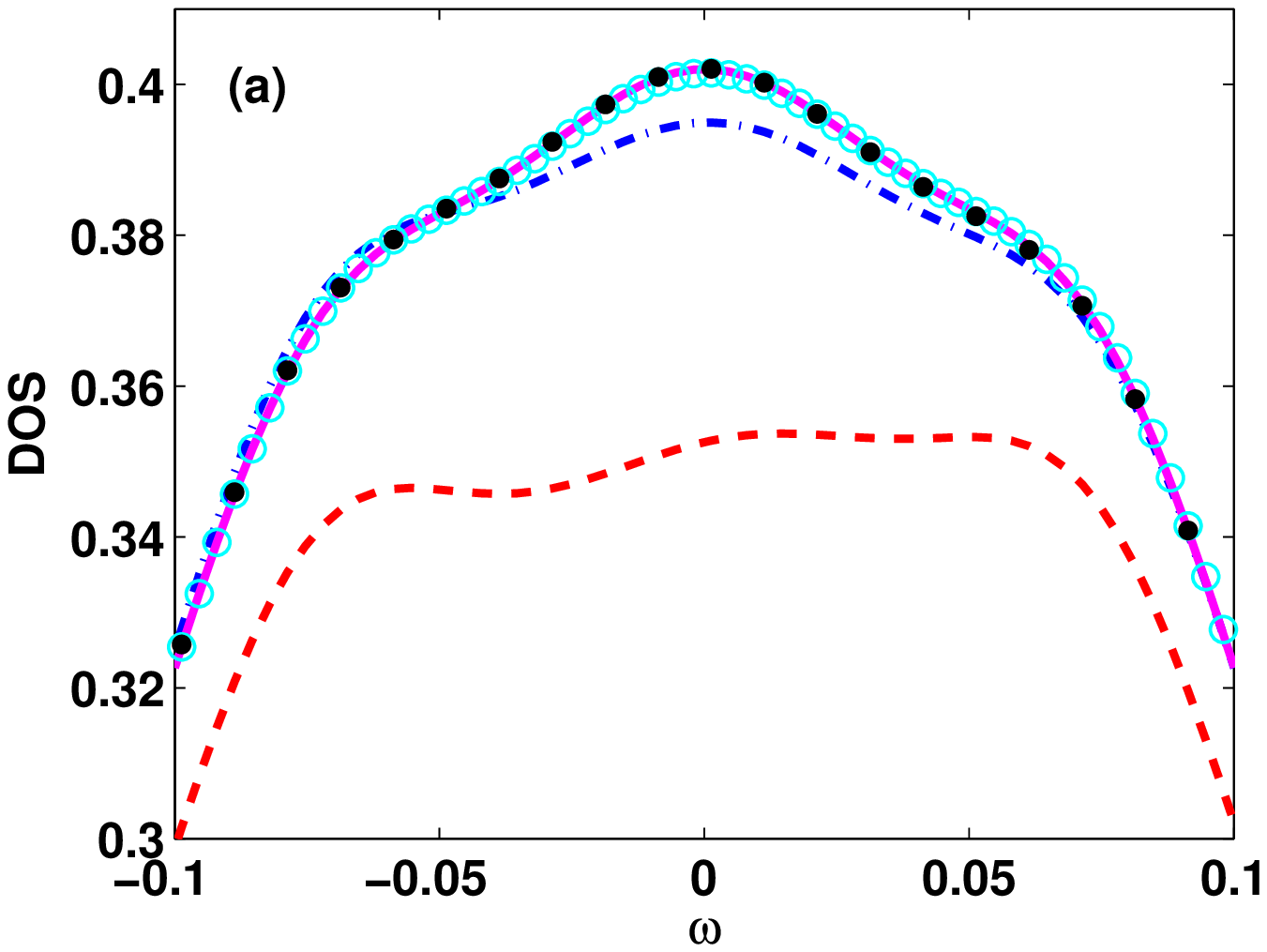}}
  \mbox{\includegraphics[scale=0.5]{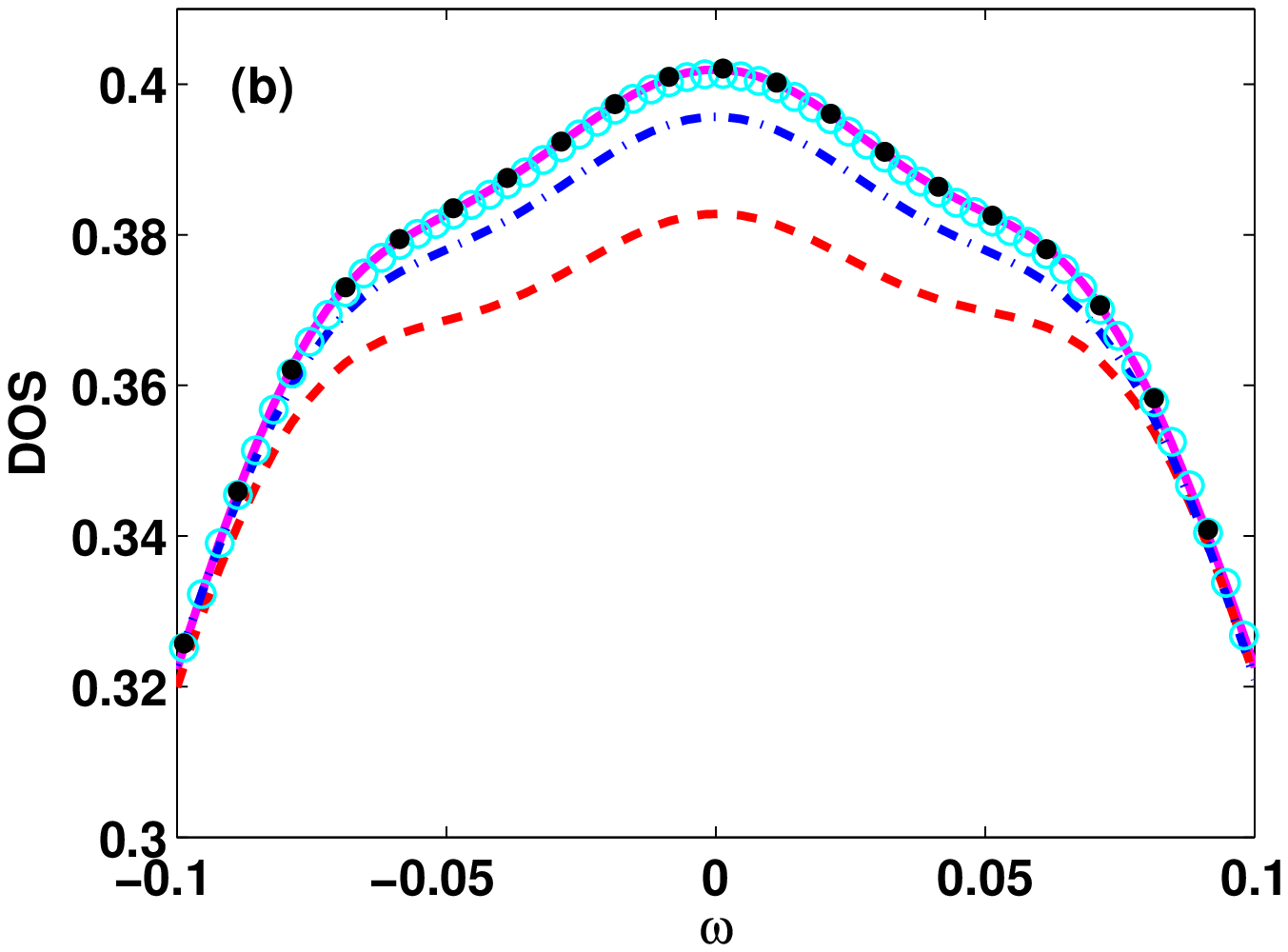}}
  \mbox{\includegraphics[scale=0.5]{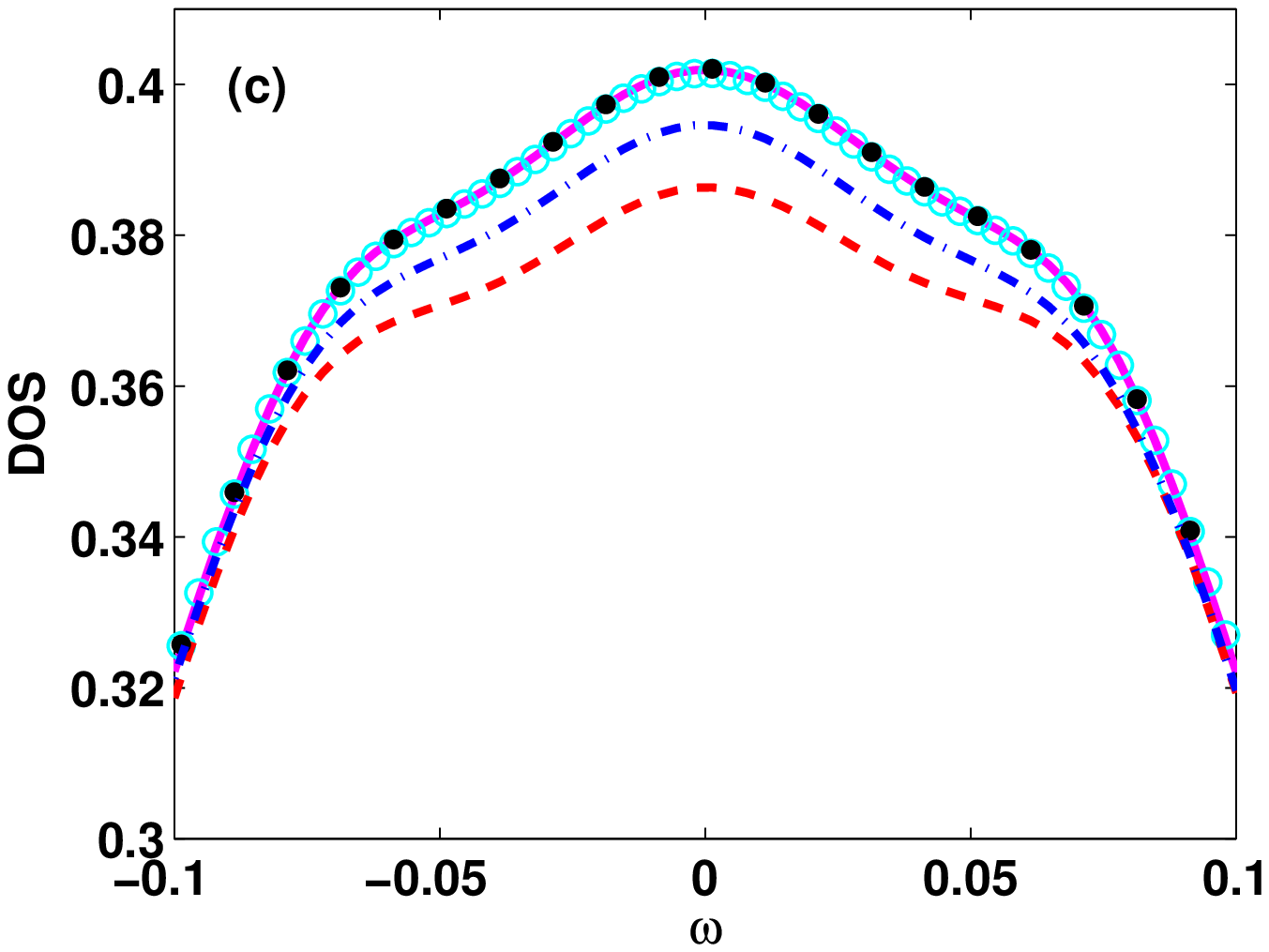}}
  \mbox{\includegraphics[scale=0.5]{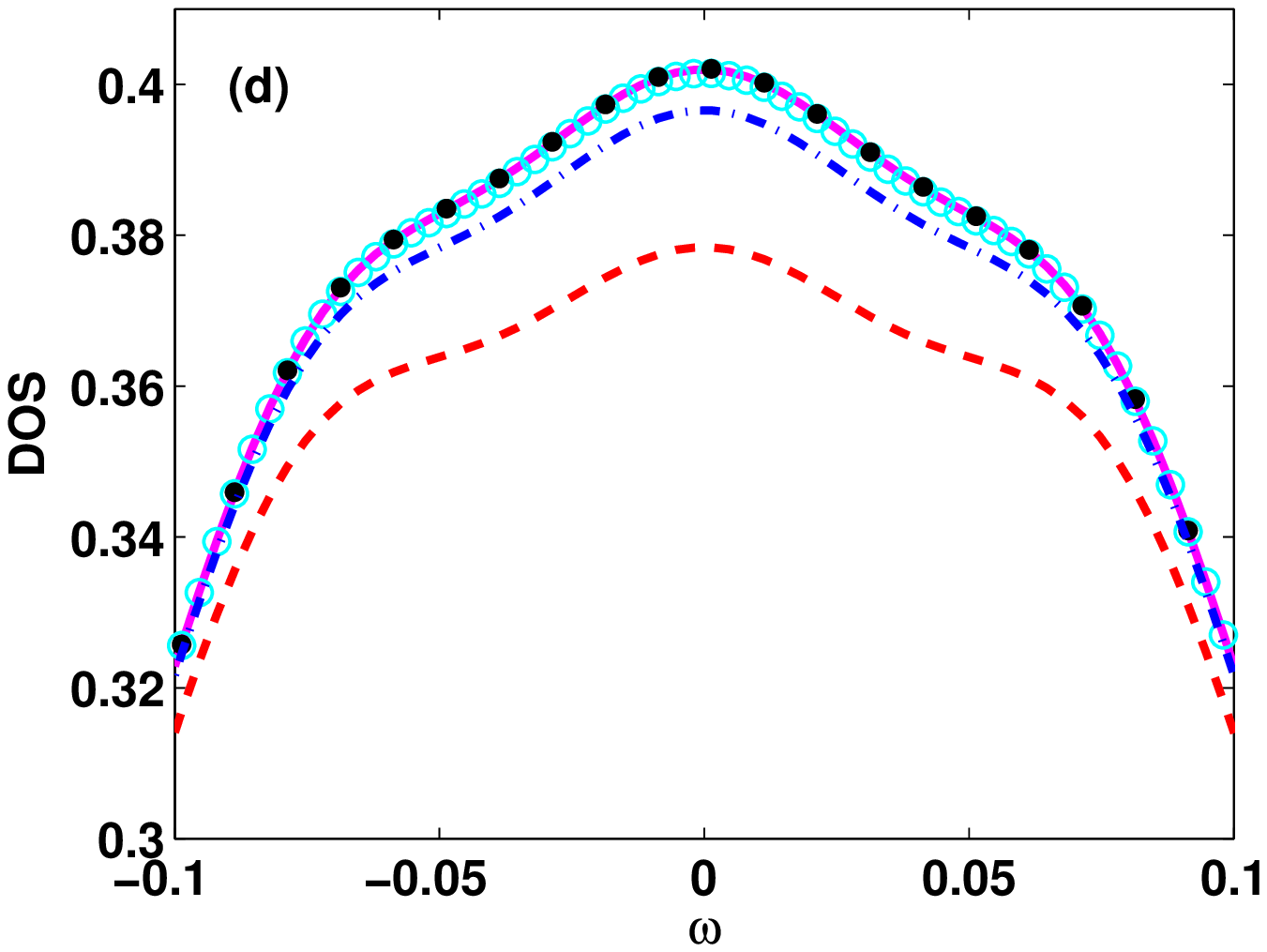}}
  \end{center}
  \caption{(Color online) machine learning prediction for the DOS around $\omega = 0$ for the different representations of the Green's function and length of the learning set. (a) Continued fraction. (b) Matsubara frequency. (c) Imaginary time. (d) Legendre polynomials. The length of the learning sets are as follows: red dashed lines (- -) denote the result for a learning set of length 500, blue dot-dashed lines (- .) denote the result for a learning set of length 1000, cyan circles (o) denote the result for a learning set of length 2000, and magenta solid line (-) denote the result for a learning set of length 3000. Dots denote the exact result.}
  \label{fig:predic_DOS_nd1_4_representations}
\end{figure*}

It is important to show how our prediction of the DOS evolves with increasing learning set length. Fig.~\ref{fig:ARD_coeffs} displays the learning set length dependence of the $ARD$. Now we look at the DOS around the Fermi level for the half-filled case. The conclusions we show for the DOS around the Fermi level are generally applicable to the high-frequency results also. One exception is the continued fraction representation where much smaller learning sets can predict the high-frequency behavior (see Fig.~\ref{fig:predic_wn_w_coeffs}). In Fig.~\ref{fig:predic_DOS_nd1_4_representations}, we show the DOS around $\omega = 0$ for different random learning set lengths and for the four representations: in Fig.~\ref{fig:predic_DOS_nd1_4_representations}(a) the continued fraction, in Fig.~\ref{fig:predic_DOS_nd1_4_representations}(b) the Matsubara frequency, in Fig.~\ref{fig:predic_DOS_nd1_4_representations}(c) imaginary time, and in Fig.~\ref{fig:predic_DOS_nd1_4_representations}(d) Legendre polynomials. The dots (.) denote the exact result, the red dashed lines (- -) denote the result for a learning set of length 500, the blue dot-dashed lines (- .) denote the result for a learning set of length 1000, the cyan circles (o) denote the result for a learning set of length 2000, and the magenta solid lines (-) denote the result for a learning set of length 3000. We can see that convergence to the correct prediction is attained before the learning set is maximum (4999).
\subsection{Prediction of the mass enhancement}\label{QI_sec_mass_enh}
\begin{figure}[tpb]
  \begin{center}
  \mbox{\includegraphics[scale=0.5]{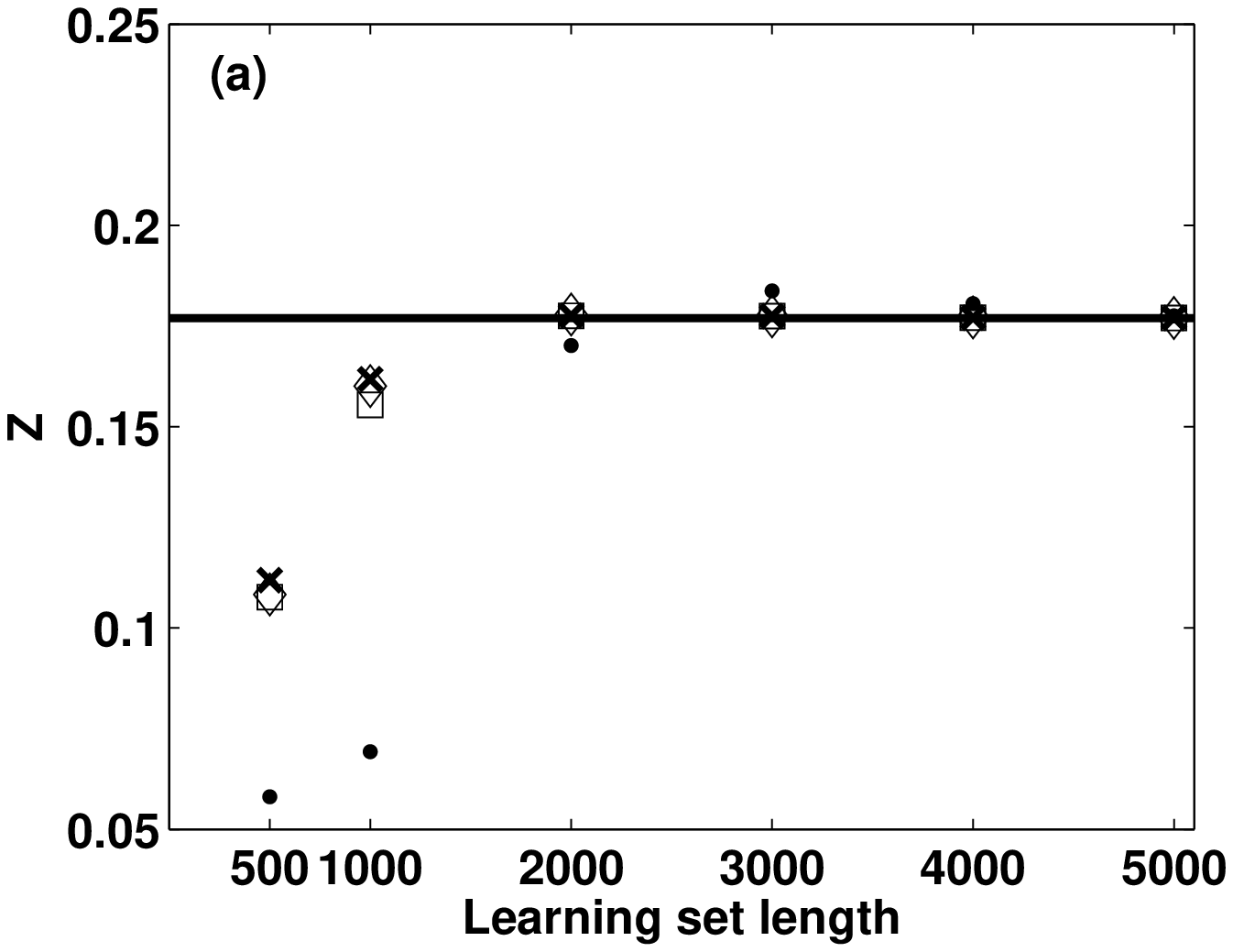}}
  \mbox{\includegraphics[scale=0.5]{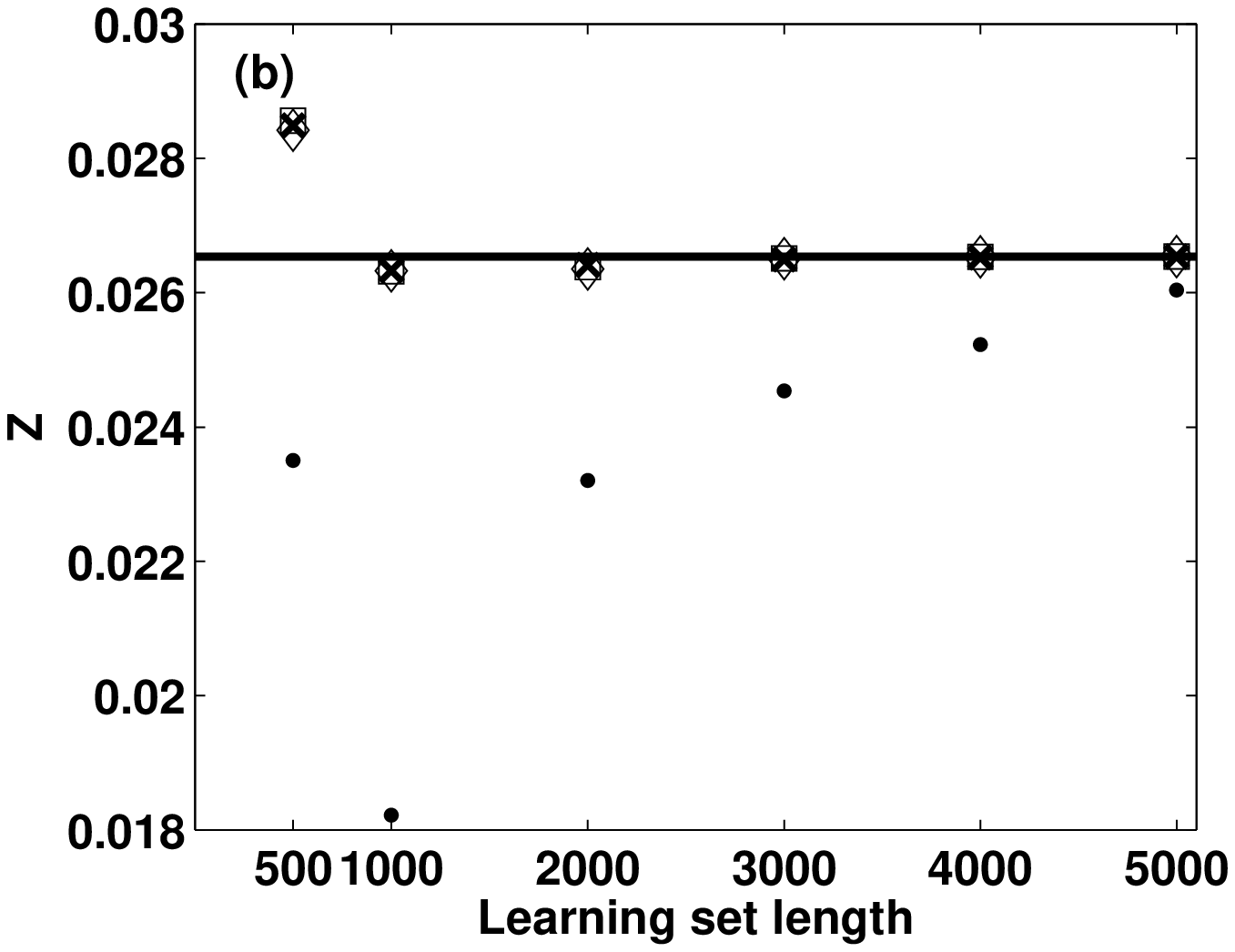}}
  \end{center}
  \caption{Quasi-particle weight $Z$ as a function of training set size (a) $U = 3.04$, $n_d = 1$, and $V = 0.5105$ and (b) $U = 4$, $n_d = 0.95$, and $V = 0.3053$. Lines denote the exact results, dots (.) denote the predictions of the continued fraction coefficients, x's (x) denote the predictions of $G(i\omega_n)$, squares ($\square$) denote the predictions of $G(\tau)$, and diamonds ($\diamondsuit$) denote the Legendre polynomial expansion.}
  \label{fig:Z}
\end{figure}

We now consider how well machine learning  predicts the renormalization factor $Z$, equivalent in this model to the mass enhancement. In addition to being an important physical property for Fermi liquids, $Z$ also acts here as a unique number that can be used to estimate the quality of the ML prediction of  low-frequency properties. The results are shown in Fig.~\ref{fig:Z}. Overall, the $Z$ predicted from ML learned continued fraction coefficients never completely converges (the black line is the exact result) and is quite inaccurate for learning set of 500 and 1000 when $n_d = 1$. Even if it never perfectly converges, the relative difference at the larger learning set is small, consistent with the reasonable visual appearance of the DOS  (Fig~\ref{fig:predic_wn_w_coeffs}). For $n_d = 0.95$ there is a peculiar concordance where the prediction is better for 500 than 1000 and 2000. This is clearly accidental and the overall shape of the DOS around $\omega = 0$ is better for 1000 and 2000. It is also important to note that for Fig.~\ref{fig:Z}-(b) the exact $Z$ is small and thus we predict very small numbers. For the other three representations, they all systematically converged to the correct answer, around a learning set of 2000 for Fig.~\ref{fig:Z}(a), which also can be seen from Fig.~\eqref{fig:predic_DOS_nd1_4_representations}, and around a learning set of 3000 for Fig.~\ref{fig:Z}-(b). Again, these minimal converging lengths are for totally random sets of examples as learning sets.
\subsection{Prediction from a minimal learning set}\label{QI_predic_min_set}
\begin{figure}[tpbh]
  \begin{center}
  \mbox{\includegraphics[scale=0.5]{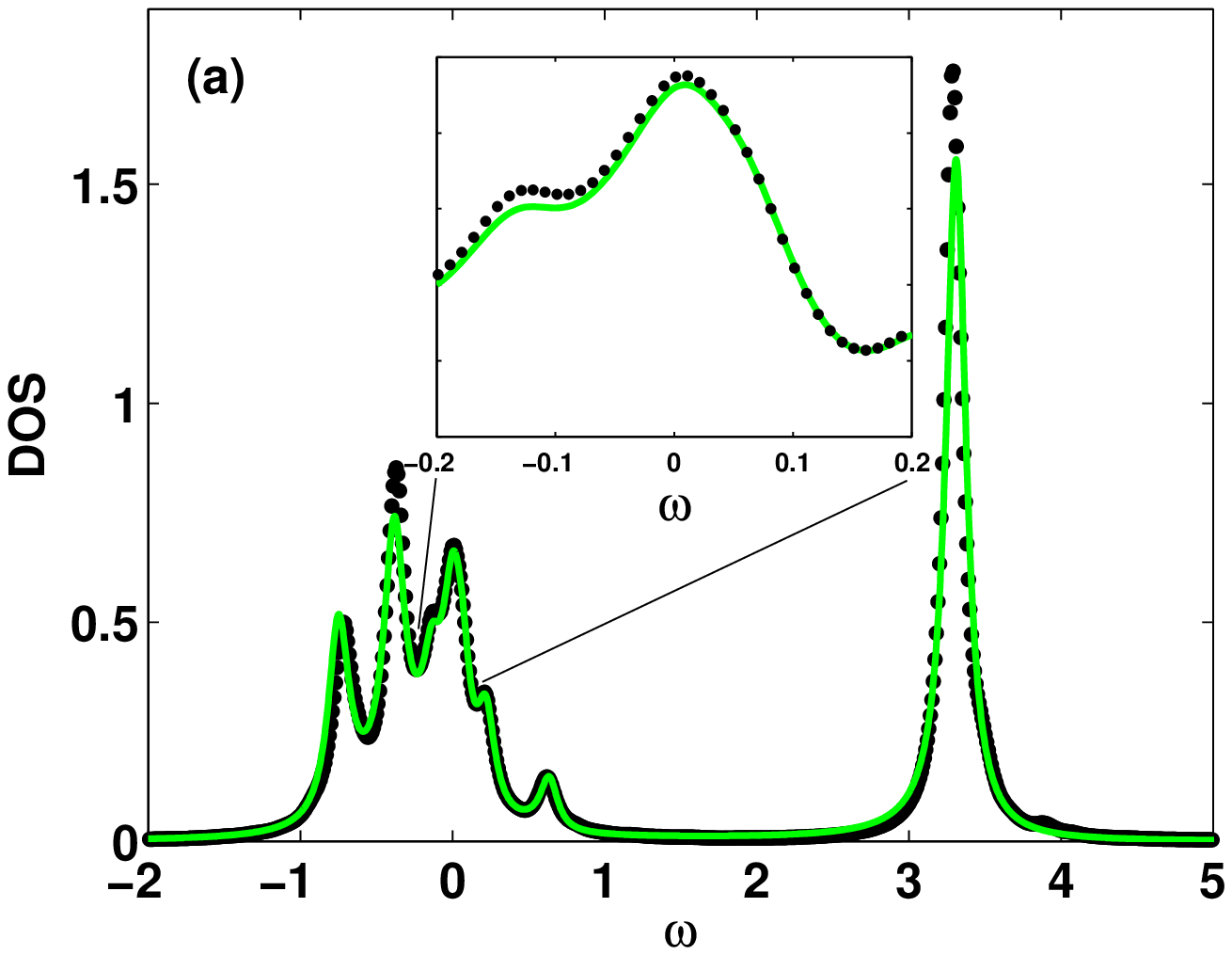}}
  \mbox{\includegraphics[scale=0.5]{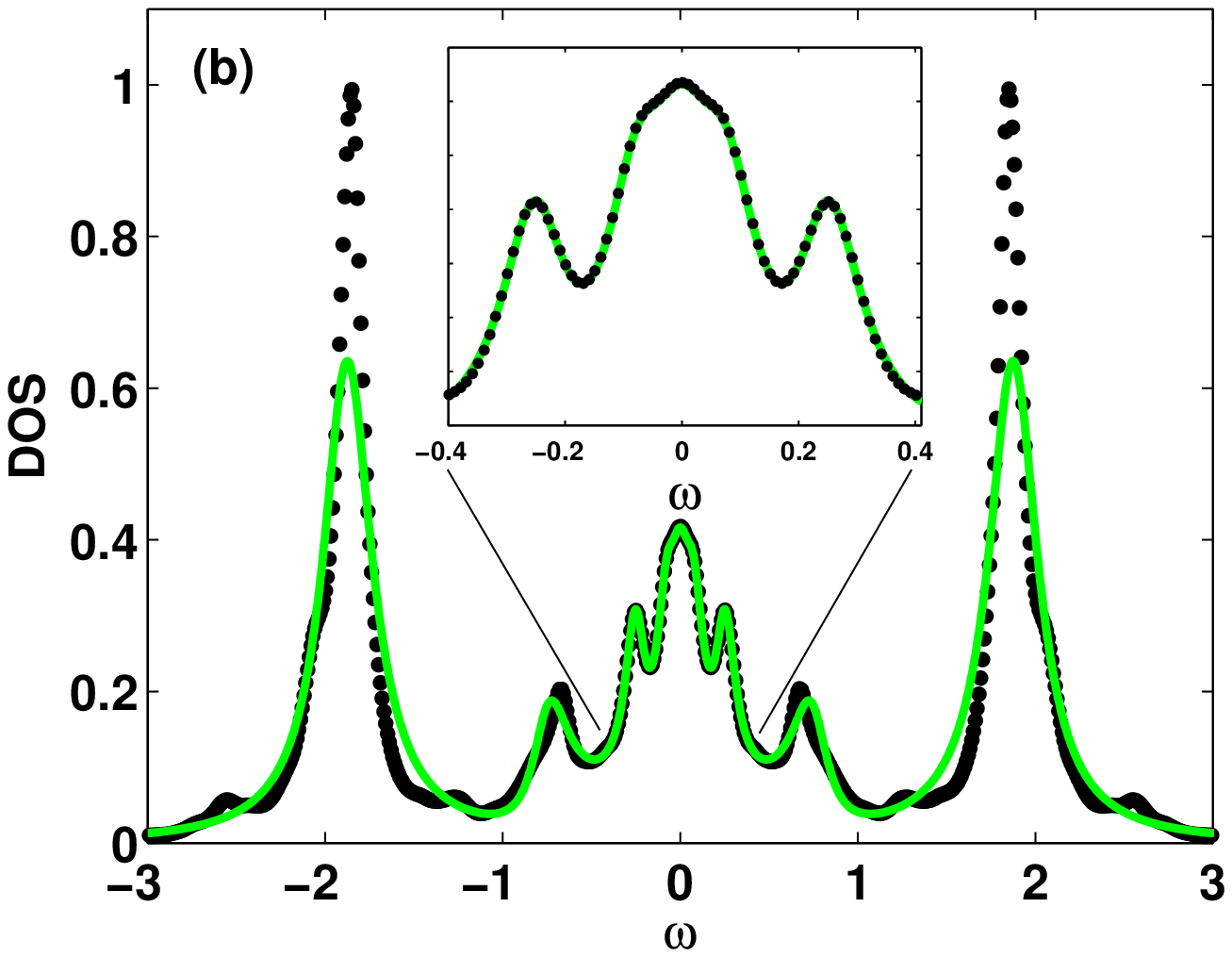}}
  \end{center}
  \caption{(Color online) machine learning prediction for the density of states using the Legendre polynomial representation: (a) $U = 3.5$, $V = 0.25$, and $n_d = 0.85$ and (b) $U = 2.9$, $V = 0.5$, and $n_d = 1$. Dots (.) denote the exact result, and green lines (-) denote the result for a learning set of length ((a)) 8  and ((b)) 4.}
  \label{fig:predic_wn_w_Gl_min}
\end{figure}

\begin{figure*}[tpbh]
  \begin{center}
  \mbox{\includegraphics[scale=0.5]{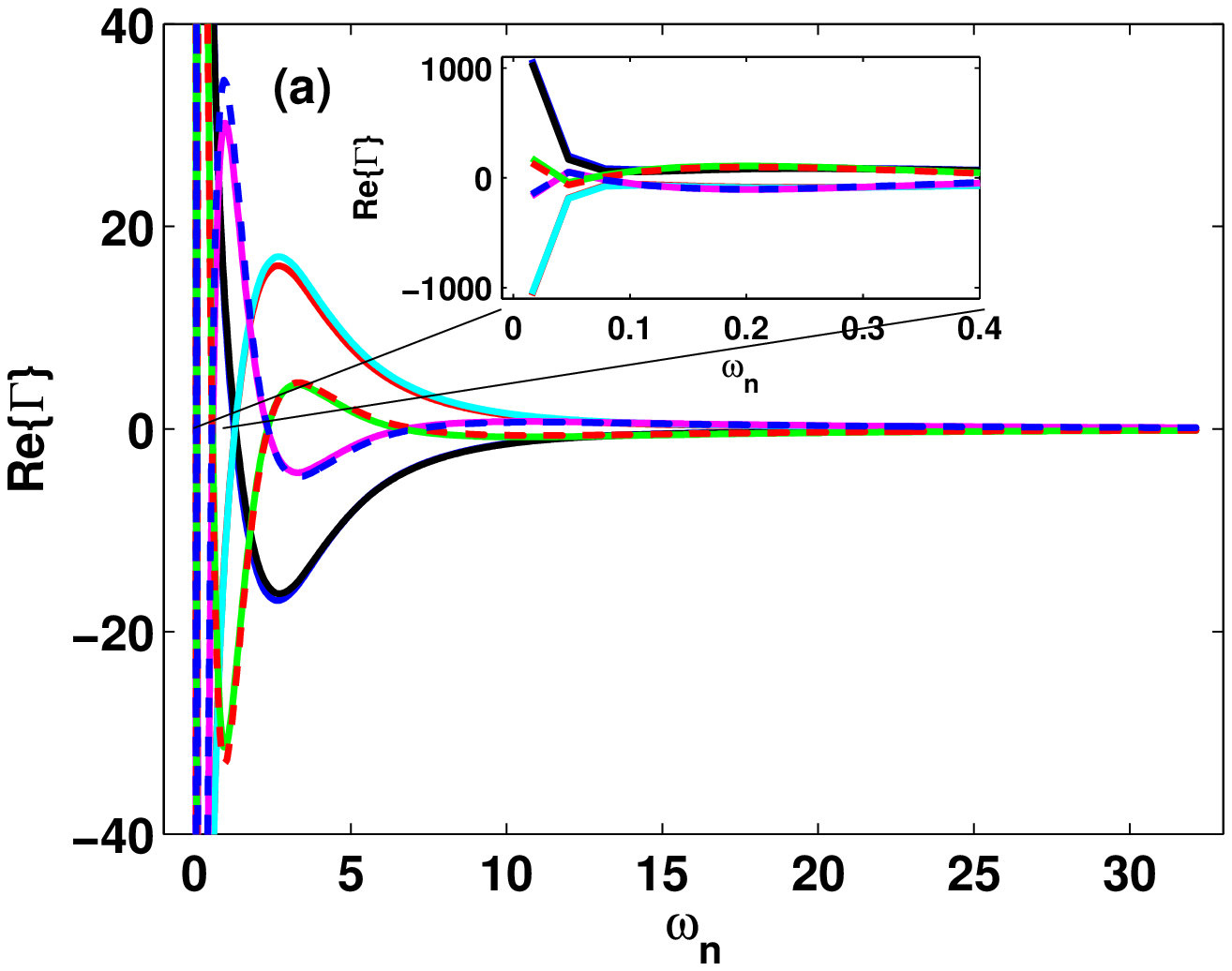}}
  \mbox{\includegraphics[scale=0.5]{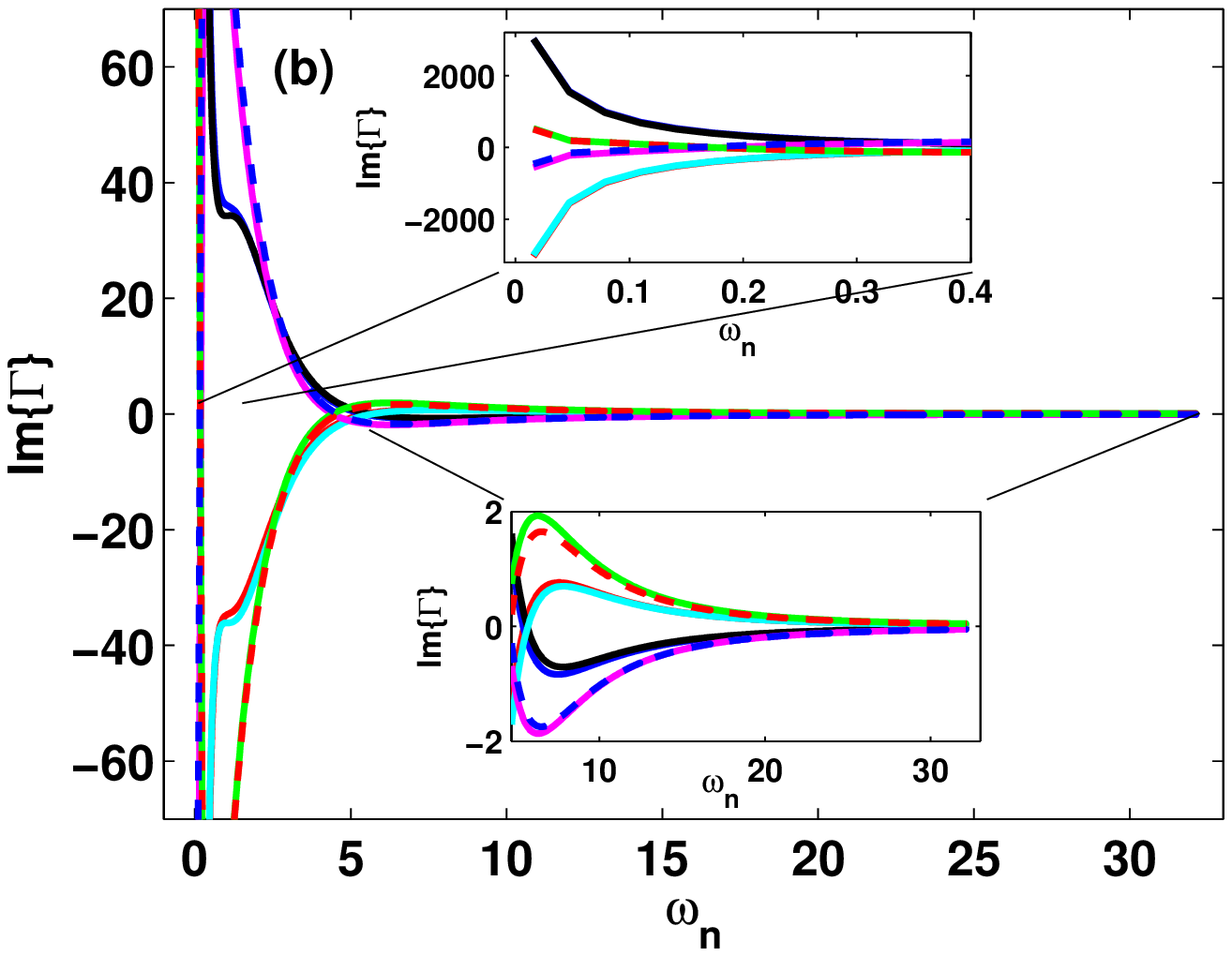}}
  \mbox{\includegraphics[scale=0.5]{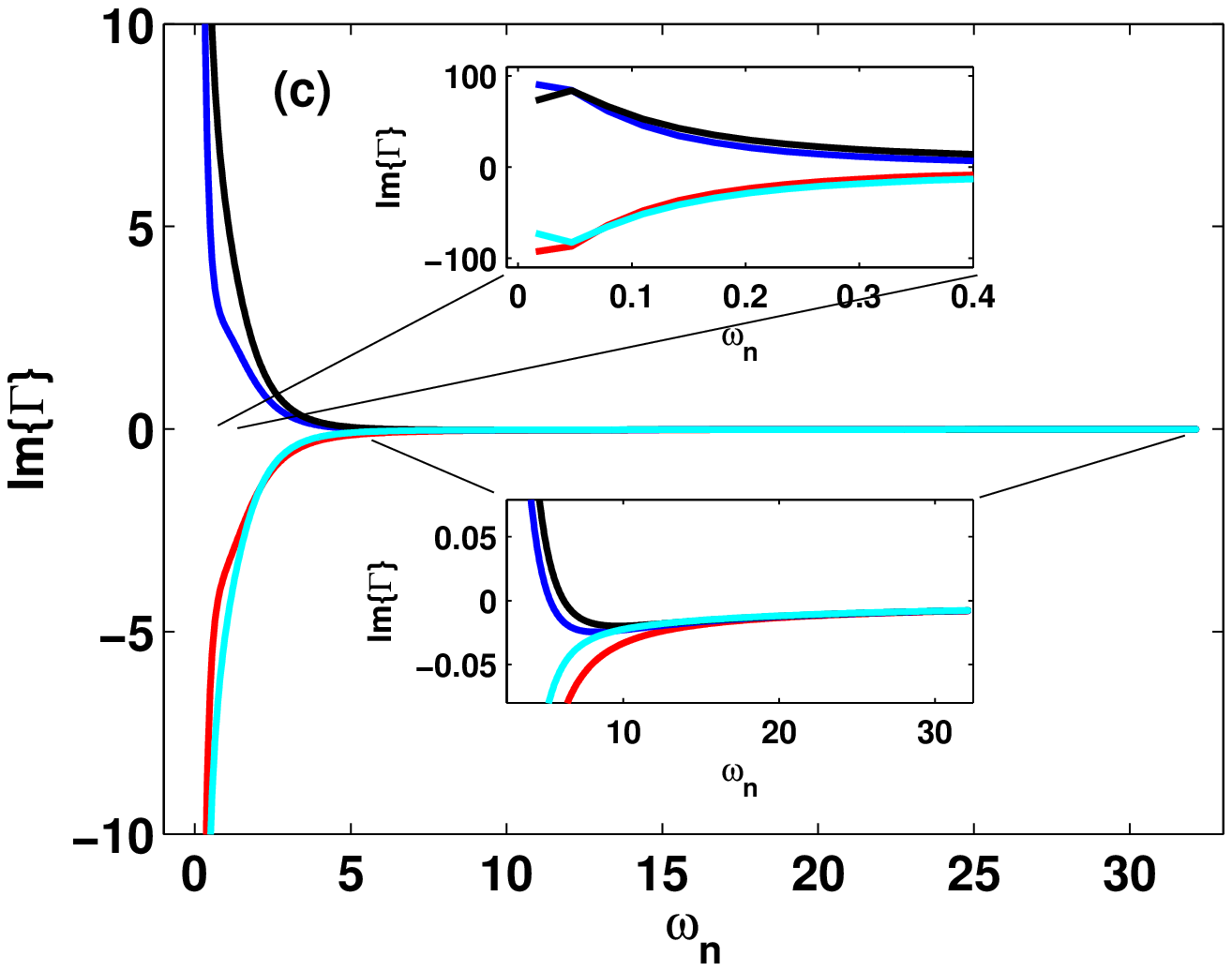}}
  \end{center}
  \caption{(Color online) Effective ML coefficient parameters $\Gamma$ for the Legendre polynomial representation as a function of $\omega_n$ for the (a) real and (b) imaginary parts $U = 3.5$,$V = 0.25$, and $n_d = 0.85$ and for the (c) imaginary part $U = 2.9$, $V = 0.5$, and $n_d = 1$. The curves correspond to the different examples in the learning set. For (a) and (b), red solid lines denote $U = 3.36$, $V = 0.24$, and $n_d = 0.80$; blue solid lines denote $U = 3.36$, $V = 0.24$, and $n_d = 0.90$; black solid lines denote $U = 3.36$, $V = 0.27$, and $n_d = 0.80$; cyan solid lines denote $U = 3.36$, $V = 0.27$, and $n_d = 0.9$; magenta solid lines denote $U = 3.52$, $V = 0.24$, and $n_d = 0.80$; green solid lines denote $U = 3.52$, $V = 0.24$, and $n_d = 0.90$; red dashed lines denote $U = 3.52$, $V = 0.27$, and $n_d = 0.80$; and blue dashed lines denote $U = 3.52$, $V = 0.27$, and $n_d = 0.90$. For (c), red solid lines denote $U = 2.88$, $V = 0.48$, and $n_d = 1$; blue solid lines denote $U = 2.88$, $V = 0.51$, and $n_d = 1$; black solid lines denote $U = 3.04$, $V = 0.48$, and $n_d = 1$; and cyan solid lines denote $U = 3.04$, $V = 0.51$, and $n_d = 1$.}
  \label{fig:alpha_Gl}
\end{figure*}

\begin{figure*}[tpbh]
  \begin{center}
  \mbox{\includegraphics[scale=0.5]{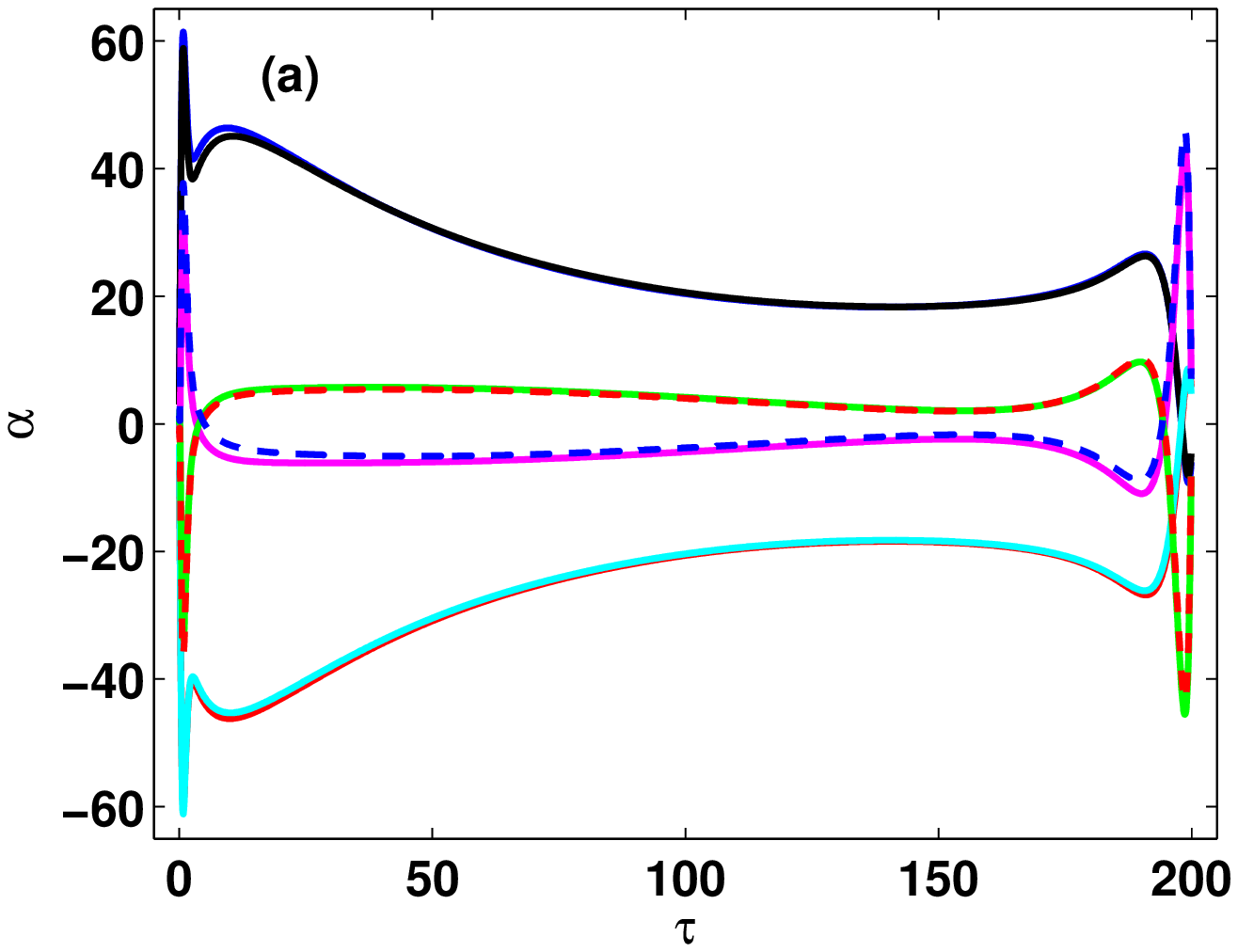}}
  \mbox{\includegraphics[scale=0.5]{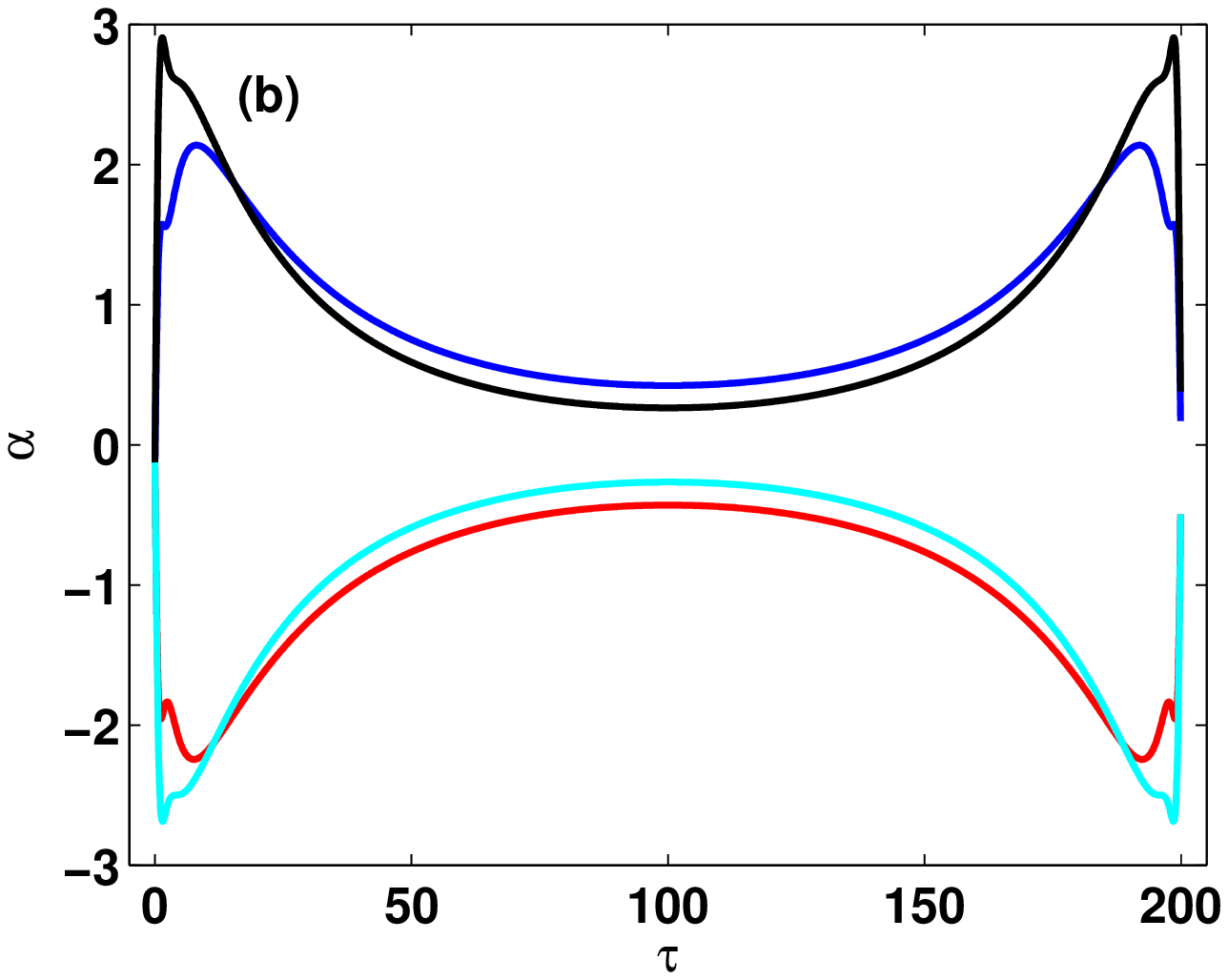}}
  \end{center}
  \caption{(Color online) Effective ML coefficients parameters $\Gamma$ for the Legendre polynomial representation as a function of $\tau$ for (a) $U = 3.5$, $V = 0.25$, and $n_d = 0.85$ and (b) $U = 2.9$, $V = 0.5$, and $n_d = 1$. The curves correspond to the different examples in the learning set. For (a), red solid lines denote $U = 3.36$, $V = 0.24$, and $n_d = 0.80$; blue solid lines denote $U = 3.36$, $V = 0.24$, and $n_d = 0.90$; black solid lines denote $U = 3.36$, $V = 0.27$, and $n_d = 0.80$; cyan solid lines denote $U = 3.36$, $V = 0.27$, and $n_d = 0.9$; magenta solid lines denote $U = 3.52$, $V = 0.24$, and $n_d = 0.80$; green solid lines denote $U = 3.52$, $V = 0.24$, and $n_d = 0.90$; red dashed lines denote $U = 3.52$, $V = 0.27$, and $n_d = 0.80$; and blue dashed lines denote $U = 3.52$, $V = 0.27$, and $n_d = 0.90$. For (c), red solid lines denote $U = 2.88$, $V = 0.48$, and $nd = 1$; blue solid lines denote $U = 2.88$, $V = 0.51$, and $n_d = 1$; black solid lines denote $U = 3.04$, $V = 0.48$, and $n_d = 1$; and cyan solid lines denote $U = 3.04$, $V = 0.51$, and $n_d = 1$.}
  \label{fig:alpha_tau}
\end{figure*}

In this section we study how to introduce a selection bias into the learning set such that a new system is predicted as accurately as possible with the smallest learning set possible. Because our database is very homogenous (dense coverage of $U$, $V$, and $n_d$) we can ask how good the learning would be if we find the members of the dataset that are closest to the $U_{new}$, $V_{new}$, and $n_{d,new}$ and form the learning set as combinations thereof. Since the descriptor has three components and, at most, the new parameters can be between two values of the database, this would give us a maximally localized learning set of minimal size 8 or less. Let us once again choose two examples for which we will do the predictions. The first example is $U = 3.5$, $V = 0.25$, and $n_d = 0.85$. The second example is $U = 2.9$, $V = 0.5$, and $n_d = 1$.\\
\\
For the first case, none of the results in the 5000 database share any of these parameters so that none of the components of the difference between the descriptor of the example and any descriptors of the database is ever zero. This means that the learning set will be of length 8. In the second example, since we look at half-filling and the database contains half-filled results, the learning set size is 4. This means that we are really looking at how our ML scheme can predict new results. We show the results in Fig.~\ref{fig:predic_wn_w_Gl_min} using the Legendre polynomial representation and we see that with a very small learning set we can predict quite precisely new results.\\
\\
This possibility of prediction with very small learning enables us to look closely at how the machine itself behaves. Indeed, what we call the machine is given by the $\alpha$ parameters of Eq.~\eqref{MLfundamental}. Only in the cases of the representations in terms of Matsubara frequencies and imaginary time does the $\alpha$ really represent the Green's function we are trying to predict itself. However, within the Legendre polynomial representation, using our approximation that the Kernel functions have fixed parameters, we can define an effective $\alpha$-like parameter for the reconstructed Green's function in Matsubara frequency from the predicted Legendre polynomials coefficients. We show the derivation in Appendix~\ref{Appen:eff_alpha}, and we obtain
\begin{equation}\label{G_wn_alpha_renom}
  G(i\omega_n,\mathbf{D})=\sum_i\Gamma_{in}K(\mathbf{D}_i,\mathbf{D}),
\end{equation}
where the effective $\alpha$, called $\Gamma$, is given by
\begin{equation}\label{Gamma}
  \Gamma_{in} = \left\{\sum_{l=0}^{l_{max}}T_{nl}\alpha_{il}\right\}.
\end{equation}
The results are shown in Fig.~\ref{fig:alpha_Gl}. The case of $U = 3.5$, $V = 0.25$, and $n_d = 0.85$ is presented in Figs.~\ref{fig:alpha_Gl}(a) and (b) while the case $U = 2.9$, $V = 0.5$, and $n_d = 1$ is presented in Fig.~\ref{fig:alpha_Gl}(c). For the half-filled example, only the imaginary part of $\Gamma$ is shown, as the real part of $G(i\omega_n)$ is zero for the particle-hole symmetric case. Despite the discrete nature of functions of  Matsubara frequency, we present the curves as continuous here.  It is also interesting to note (not shown) that the curves for $\alpha$ obtained from  the Matsubara frequency representation correspond exactly to the curves of $\Gamma$ in Fig.~\ref{fig:alpha_Gl} as expected. One very important result is that $\Gamma$ is, from quite a low frequency, a smooth function of $\omega_n$. This opens up the possibility of  using interpolation from $\Gamma$ curves as a another way to perform very efficient ML.
For completeness, the other representation that directly learns $G$ is the imaginary time one and thus $\alpha (\tau)$ is really a representation of $G(\tau)$ in the ML space. We thus show the $\alpha (\tau)$ curve for both examples in Fig.~\ref{fig:alpha_tau}. Once again the curves are  smooth and in this case fundamentally continuous.
\section{Summary and Conclusion}\label{sum_concl}
We have proposed a machine learning  scheme to learn the electron Green's function. Our method should apply to any other correlation function of interest. We have reduced the problem of learning a function of a single variable into one of learning a relatively small set of independents numbers, either direct slices of the function or coefficients used to define it. For the Green's function of the single-site Anderson impurity model, we have tested four different representations of $G$: the continued fraction, the Matsubara frequency, imaginary time and Legendre polynomial expansion. Directly learning the function in imaginary time is a well-defined operation as long as the learning set is not too small and random. However, replacing the direct function in imaginary time $\tau$ by its Legendre polynomial expansion is clearly superior because the need to learn only a small number of coefficients improves the accuracy. This way is even more promising for the context of learning correlation functions for real materials as new powerful CTQMC algorithms directly measure these coefficients\cite{Boehnke_Legendre}. We also observe that the Matsubara frequency  representation may be problematic  for more general DMFT calculations. In the AIM studied here, the system is always a metal and thus the dataset of solved problems in $\omega_n$ is very homogeneous. In the general case of the DMFT, where the AIM serves as an intermediate problem, there is an interaction driven metal to insulator transition. In Matsubara frequency, the Green's function at lower frequencies changes qualitatively from the metallic to the insulating phase. This creates a dataset much less homogeneous at low $\omega_n$. This is why, even if this representation yields accurate predictions in the present study, we do not think that it can be efficiently used in general for ML+DMFT. Contrary to $G(i\omega_n)$, $G(\tau)$ changes much less drastically from a metal to an insulator. Therefore, this representation should perform much better for ML. Finally, as we already discussed, the continued fraction representation can only be really used when the AIM is solved via ED and thus is a very limited representation.\\
\\
Before handling real materials, learning DMFT for model Hamiltonians is the next logical step. Concomitant to this, the next logical step for ML of a function itself is to look for a scheme where learned numbers are not considered to be totally independent. This is highly nontrivial and perhaps could be achieved by adding appropriating new constraints to the minimization problem inherent in Kernel ridge regression of Eq.~\eqref{cost_set}. For example, constraints concerning moments could perhaps be integrated in the ML scheme for learning functions that have a spectral representation.
\begin{acknowledgments}
This research was supported by the Office of Science of the U.S. Department of Energy under Subcontract No. 3F-3138. L-F.A. thanks Ara Go for numerous discussions on implementing an exact diagonalization code for the Anderson impurity model. We thank Peter B. Littlewood for discussions and critical reading of the manuscript. This research used resources of the Argonne Leadership Computing Facility at Argonne National Laboratory, which is supported by the Office of Science of the U.S.~DOE under Contract No.DE-AC02-06CH11357. O.A.v.L acknowledges funding from the Swiss National Science Foundation Grant No.~PPOOP2\_ 138932.
\end{acknowledgments}
\appendix
\section{Kernel ridge regression}\label{ML_appen}
In kernel ridge regression, the coefficients of the expansion in the kernel space of Eq.~\eqref{MLfundamental} are found by minimization with respect to $\alpha$ of the cost function (Eq.~\eqref{cost_set})
\begin{eqnarray}\label{cost_set_app}
  C_m &=& \sum_l\left( g_m(\mathbf{D}_l) - f_m(\mathbf{D}_l) \right)^2
  \\
  &&+ \lambda_m\sum_{l,p}\alpha_{lm}K_m(\mathbf{D}_l,\mathbf{D}_p)\alpha_{pm}.
  \nonumber
\end{eqnarray}
The $\alpha_{lm}$ are determined as the solution of the set of equations $\d{\delta C_m}{\delta\alpha_{qm}}=0$. By using the definition of $g_m(\mathbf{D})$ given by Eq.~\eqref{MLfundamental} and also the fact that the kernel is a symmetric function, we obtain
\begin{eqnarray}\label{deriv_inter_app}
&&\sum_l\Bigg[ \left(\sum_p\alpha_{pm}K_m(\mathbf{D}_p,\mathbf{D}_l)\right) - f_m(\mathbf{D}_l) + \lambda_m\alpha_{lm} \Bigg]
\nonumber \\
&&\times K_m(\mathbf{D}_l,\mathbf{D}_q)= 0
\end{eqnarray}
A sufficient condition for Eq.~\eqref{deriv_inter_app} to hold is that the quantity in the square brackets vanishes, i.e.
\begin{equation}\label{deriv_inter_app_1}
  \left(\sum_p\alpha_{pm}K_m(\mathbf{D}_p,\mathbf{D}_l)\right) - f_m(\mathbf{D}_l) + \lambda_m\alpha_{lm} = 0.
\end{equation}
In Eq.~\eqref{deriv_inter_app_1}, $m$ is a dummy index. For fixed $m$, we can represent $\alpha$ and $f$ as vectors $\boldsymbol{\alpha}_m$ and $\boldsymbol{f}_m$ of length given by the number of examples $N_L$ and the kernel as a $N_L\times N_L$ matrix $\overline{\overline{ \boldsymbol{K}}}_m$ so that
\begin{equation}\label{alphadef_app}
\boldsymbol{\alpha}_m = \left( \overline{\overline{ \boldsymbol{K}}}_m + \lambda_m\overline{\overline{ \boldsymbol{I}}} \right)^{-1}\boldsymbol{f}_m,
\end{equation}
with $\overline{\overline{ \boldsymbol{I}}}$ the identity matrix.
 \section{Exact Diagonalization}\label{ED_appen}
In this appendix we describe what is meant by exact diagonalization (ED) in the case of the Anderson impurity Hamiltonian. Here, Eq.~\eqref{H_AIM} is the target Hamiltonian $H_{target}$. To solve the problem by ED, we map $H_{target}$ to a new Hamiltonian where the effect of the continuous bath is approximated by a few poles and weights. This gives a Hamiltonian with a finite number of site $N_s = 1+N_b$ given by
\begin{equation}\label{H_ED_appen}
\begin{split}
  H = &\sum_{\sigma}\varepsilon_dd_{\sigma}^{\dagger}d_{\sigma} + Ud_{\uparrow}^{\dagger}d_{\uparrow}d_{\downarrow}^{\dagger}d_{\downarrow} + \sum_{l=1,\sigma}^{N_b}\varepsilon_l c_{l\sigma}^{\dagger}c_{l\sigma}\\ &+ \sum_{l=1,\sigma}^{N_b}V_l\left( d_{\sigma}^{\dagger}c_{l\sigma} + c_{l\sigma}^{\dagger}d_{\sigma} \right),
\end{split}
\end{equation}
where the $V_l$ and $\varepsilon_l$ are chosen to reproduce as much as possible the effect of Eq.~\eqref{Deltadef}. We have now a finite size Hamiltonian with a Hilbert space of size $4^{N_s}$ that can be solved using matrices diagonalization techniques. We use the usual Lanczos approach\cite{Dagotto}. Of course, by using a finite bath, it is impossible to reproduce the continuous case to perfect match. Only in the case where the number of bath sites $N_b$ was equal to infinity could one recover the continuous case exactly. Therefore, the idea proposed by Caffarel and Krauth\cite{ED_DMFT} is to define a distance function $d$ from the Matsubara axis representation. Note that the Matsubara axis representation of a $T=0$ problem requires the use of a fictitious temperature, which we choose as $\beta = 1/T = 200$. Here, we use the inverse of the noninteracting Green's function to define the distance function
\begin{equation}\label{distance_appen}
  d = \d{1}{N_{max}+1}\sum_{n=0}^{N_{max}}\d{\left| G_0^{-1}(i\omega_n)-G_0^{-1,Ns}(i\omega_n)  \right|^2}{\omega_n}.
\end{equation}
$N_{max}$ is the maximum number of frequencies used to define $d$. It is important that $\omega_{N_{max}} \gg max (\varepsilon_l)$. We typically use $N_{max}=400$. Since the bath is fixed with a half bandwidth of 1, $max(|\varepsilon_l|) \leq 1$. At $\beta = 200$ such an energy corresponds to approximately $n = 31$ and thus $N_{max} = 400$ is large enough. Finally, $G_0^{-1,Ns}(i\omega_n)$ is the inverse of the noninteracting Green's function of the Hamiltonian of Eq.~\eqref{H_ED_appen} and is written as
\begin{equation}\label{G0_ED_appen}
  G_0^{-1,Ns}(z) = z - \varepsilon_d - \underbrace{\sum_{l=1}^{N_b}\d{V_l^2}{z-\varepsilon_l}}_{= \Delta_{N_b}}.
\end{equation}
We specify the set of parameters $\{V_l,\varepsilon_l\}$ as those that minimize $d$. This is a problem of unconstrained optimization in several variables.
\subsection{Representation of the Green's function}
For our $H_{target}$ (Eq.~\eqref{H_ED_appen}) with hybridization fixed, the ground state is nondegenerate and always in the sector where $N=N_s$ and $S_z = 0$. Once the energy $E_{GS}$ and many-body wave function $|GS\rangle$ are obtained, the Green's function can be calculated from the electron and hole part.
\begin{equation}\label{G_def_appen}
\begin{split}
  G_{\sigma}(z) = &\langle GS|d_{\sigma}\d{1}{z+E_{GS}-H}d_{\sigma}^{\dagger}|GS\rangle\\ &+ \langle GS|d_{\sigma}^{\dagger}\d{1}{z-E_{GS}-H}d_{\sigma}|GS\rangle.
\end{split}
\end{equation}
It is most convenient to represent $G$ as a continued fraction
\begin{equation}\label{G_cont_frac_appen}
\begin{split}
  G_{\sigma}(z) = &\d{\langle GS|d_{\sigma}d_{\sigma}^{\dagger}|GS\rangle}{z+E_{GS}-a_0^>-\d{b_1^{>2}}{z+E_{GS}-a_1^>-\d{b_2^{>2}}{z+E_{GS}-a_2^> - \ddots}}}\\ &+ \d{\langle GS|d_{\sigma}^{\dagger}d_{\sigma}|GS\rangle}{z-E_{GS}-a_0^<-\d{b_1^{<2}}{z-E_{GS}-a_1^<-\d{b_2^{<2}}{z-E_{GS}-a_2^< - \ddots}}}.
\end{split}
\end{equation}
\\
\\
A second Lanczos procedure can be used to obtained the coefficients. The algorithm to calculate these coefficients from the ground state is as follows\cite{Dagotto}:\\
\\
\begin{itemize}
\item
We construct a starting vector $|f_0\rangle = d_{\sigma}^{\dagger}|GS\rangle$
\item
We construct the so-called Lanczos space with the recursion relation of orthogonal vectors
\begin{equation}\label{recu_Lan_appen}
  |f_{n+1}\rangle = H|f_n\rangle - a_n^>|f_n\rangle - b_n^{>2}|f_{n-1}\rangle.
\end{equation}
\item
It is easy to obtain the coefficients as
\
\begin{equation}\label{an_appen}
  a_n^> = \d{\langle f_n|H|f_n\rangle}{\langle f_n|f_n\rangle}
\end{equation}
and
\begin{equation}\label{bn_appen}
  b_n^{>2} = \d{\langle f_n|f_n\rangle}{\langle f_{n-1}|f_{n-1}\rangle},
\end{equation}
with $b_0^{>2} = 0$.
\end{itemize}
The equivalent can be done for the $a_n^<$ and $b_n^{<2}$. The $|f_n\rangle$ are orthogonal by construction if evaluated exactly. With double precision floating point arithmetic and a recursion relation (Eq.~\eqref{recu_Lan_appen}) that only forces three vectors to be orthogonal, at some point (typically after 50) the orthogonality will be lost. For the Green's function, it does not matter much. If necessary, there are slightly more complicated algorithms that do partial orthogonalization to keep a good orthogonality of our set of vectors\cite{Qiao}. Note that for proper numerical calculation Eqs.~\eqref{recu_Lan_appen},\eqref{an_appen} and \eqref{bn_appen} must be modified so that the states in Eq.~\eqref{recu_Lan_appen} are normalized.
\section{Legendre polynomial expansion}\label{Legendre_appen}
The Legendre polynomials $P_k(x)$ are defined on an interval $x$ $\epsilon$ $[-1,1]$ and thus an arbritrary function $f(x)$ in $-1 \leq x \leq 1$ can be written as
\begin{equation}\label{Leg_exp_appen}
  f(x) = \sum_{k = 0}^{\infty} a_kP_k(x).
\end{equation}
The coefficients $a_k$ are found by multiplying both sides of Eq.~\eqref{Leg_exp_appen} by $P_l(x)$ and then integrating over $x$. The orthogonality relation for the Legendre polynomials is used, $\int_{-1}^{1}dx P_l(x)P_k(x) = \d{2}{2l+1}\delta_{lk}$.
\begin{equation}\label{coeff_Leg_appen}
  \int_{-1}^1dxP_l(x)f(x) = \sum_{k=0}^{\infty} a_k\int_{-1}^1dxP_l(x)P_k(x) = a_l\d{2}{2l+1}
\end{equation}
Hence,
\begin{equation}\label{coeff_Leg_1_appen}
  a_n = \d{2l+1}{2}\int_{-1}^1dxP_l(x)f(x).
\end{equation}
We define $a_l \equiv \d{\sqrt{2l+1}}{\beta}G_l$ to be consistent with Ref.[\onlinecite{Boehnke_Legendre}]. Finally, we can also change the integration variable and therefore get the result of Eq.~\eqref{coeff_Leg_Gtau} and the expansion of Eq.~\eqref{Leg_exp_Gtau}.
\section{Effective $\alpha$ in the Machine Learning of the Legendre polynomial expansion}\label{Appen:eff_alpha}
In the Legendre polynomial expansion, the Green's function on the Matsubara axis is given by Eq.~\eqref{Leg_exp_Giwn}
\begin{equation}\label{G_wn_app}
  G(i\omega_n) = \sum_{l=0}^{\infty}T_{nl}G_l.
\end{equation}
In this approach, the ML procedure is for the coefficients $G_l$ and not $G(i\omega_n)$ itself. Therefore, using Eq~\eqref{MLfundamental} we may write
\begin{equation}\label{G_Leg_ML_app}
  G_l = \sum_i\alpha_{il}K_l(\mathbf{D}_i,\mathbf{D}).
\end{equation}
Therefore, putting Eq.~\eqref{G_Leg_ML_app} in Eq.~\eqref{G_wn_app} we obtain
\begin{equation}\label{G_wn_ML_app}
\begin{split}
  G(i\omega_n,\mathbf{D}) &= \sum_{l=0}^{l_{max}}T_{nl}\sum_i\alpha_{il}K_l(\mathbf{D}_i,\mathbf{D})\\
  &= \sum_i\sum_{l=0}^{l_{max}}T_{nl}\alpha_{il}K_l(\mathbf{D}_i,\mathbf{D}).
\end{split}
\end{equation}
In the most general case, this is as far as we can go. However, in our approach, we have chosen to consider that the parameters of $K$ are such that $K$ is independent of $l$, i.e., $K_l(\mathbf{D}_i,\mathbf{D})\rightarrow K(\mathbf{D}_i,\mathbf{D})$. This enables us to factor out $K$ in Eq.\eqref{G_wn_ML_app} and thus obtain
\begin{equation}\label{G_wn_alpha_renom_app}
\begin{split}
  G(i\omega_n,\mathbf{D}) &= \sum_i\left\{\sum_{l=0}^{l_{max}}T_{nl}\alpha_{il}\right\}K(\mathbf{D}_i,\mathbf{D})\\
  &\equiv \sum_i\Gamma_{in}K(\mathbf{D}_i,\mathbf{D}).
\end{split}
\end{equation}
We therefore obtain in this approximation a renormalized $\alpha$ parameter in the ML expansion. This can be compared with the $\alpha$ obtained from the ML directly on $G(i\omega_n)$. This is not possible for the other two representations since the reconstruction of $G(i\omega_n)$ is done using highly nonlinear relations (Fourier transform and the continued fraction).

\end{document}